\documentclass[aps,amsmath,amssymb,nofootinbib]{revtex4}

\usepackage{graphicx}% Include figure files
\usepackage{dcolumn}% Align table columns on decimal point
\usepackage{bm}% bold math

\usepackage{tikz}
\newcommand*\circled[1]{\tikz[baseline=(char.base)]{
  \node[shape=circle,draw,inner sep=1pt] (char) {#1};}}

\def\cm{{\rm cm}}
\def\crit{{\rm crit}}
\def\eff{{\rm eff}}
\def\emm{{\rm EM}}
\def\g{{\rm g}}
\def\km{{\rm km}}

\def\Mpc{{\rm Mpc}}

\def\s{{\rm s}}

\begin{document}

\title{Electrodynamics on Cosmological Scales}

\author{Li-Xin Li}
\email{lxl@pku.edu.cn}
\affiliation{Kavli Institute for Astronomy and Astrophysics, Peking University, Beijing 100871, P. R. China}

\date{\today}

\begin{abstract}
Maxwell's equations cannot describe a homogeneous and isotropic universe with a uniformly distributed net charge, because the electromagnetic field tensor in such a universe must be vanishing everywhere. For a closed universe with a nonzero net charge, Maxwell's equations always fail regardless of the spacetime symmetry and the charge distribution. The two paradoxes indicate that Maxwell's equations need be modified to be applicable to the universe as a whole. We consider two types of modified Maxwell equations, both can address the paradoxes. One is the Proca-type equation which contains a photon mass term. This type of electromagnetic field equations can naturally arise from spontaneous symmetry breaking and the Higgs mechanism in quantum field theory, where photons acquire a mass by eating massless Goldstone bosons. However, photons loose their mass again when the symmetry is restored, and the paradoxes reappear. The other type of modified Maxwell equations, which are more attractive in our opinions, contain a term with the electromagnetic field potential vector coupled to the spacetime curvature tensor. This type of electromagnetic field equations do not introduce a new dimensional parameter and return to Maxwell's equations in a flat or Ricci-flat spacetime. We show that the curvature-coupled term can naturally arise from the ambiguity in extending Maxwell's equations from a flat spacetime to a curved spacetime through the ``minimal substitution rule''. Some consequences of the modified Maxwell equations are investigated. The results show that for reasonable parameters the modification does not affect existing experiments and observations. However, we argue that, the field equations with a curvature-coupled term can be testable in astrophysical environments where the mass density is high or the gravity of electromagnetic radiations plays a dominant role in dynamics, e.g., the interior of neutron stars and the early universe.

\vspace{0.4cm}
{\bf Keywords} Classical general relativity. Maxwell's equations. Classical fields in curved spacetime. Cosmology
\end{abstract}

\maketitle

\section{Introduction}
\label{intro}

It has long been known that when some fundamental laws of physics that have been well established in labs are applied to the universe as a whole troubles or paradoxes may arise. A famous example that perfectly illustrates the above claim is the application of the second law in thermodynamics to the entire universe, which has led people to propose the idea of heat death for the ultimate fate of the universe.  The heat death hypothesis states that, when time lasts long enough, the universe will reach a thermodynamic equilibrium (and hence maximum entropy) state with absolutely uniform temperature and energy, therefore any work process and ordered movement sustained by energy will not be possible any more. It was first proposed by Thomson (Lord Kelvin) and then popularized by Helmholtz and Rankine \cite{tho62,smi89}. However, after Thomson, Planck has put doubt about the heat death by criticizing the definition for the entropy of the universe and has pointed out that such a definition has no meaning (\cite{pla10}; see also \cite{lan61,gra08} for comments on the concept of the entropy of the universe).

Modern investigations on cosmology and gravity have put more doubts on the heat death. It is well known that structure formation and evolution in the universe is dominantly governed by the law of gravity. The heat capacity of a gravitationally bound system is negative, so that the temperature of the system increases as it looses energy \cite{bin87}. Hence, it appears that the role of gravity is to keep the universe out of thermodynamic equilibrium \cite{smo14}. Einstein's theory of general relativity has led people to believe that the ultimate fate of a massive gravitationally bound system is the formation of black holes. However, studies on quantum properties of black holes have revealed that black holes are thermodynamic objects and contain a huge amount of entropy proportional to the area of their event horizons \cite{bek73,bar73}, and black holes evaporate by emitting thermal radiation \cite{haw73,haw75a}. For recent discussions on the entropy and structure formation in the universe based on the concept of gravitational entropy, please refer to \cite{cli13,sus15}. Although the problem of heat death has not been ultimately addressed yet, the interplay of general relativity and quantum theory has refreshed our understanding about the concept of entropy and may provide a solution.

Another historical example for the inconsistency of a fundamental law of physics with a global universe is the so-called Neumann-Seeliger paradox \cite{nor99}. In the 1800s, \citet{neu74} and \citet{see95} independently found that when Newton's inverse-square law of gravity was applied to an infinite and static universe with a uniform distribution of matter, one would get that the gravitational field at any position in space had a magnitude of infinity. The inconsistency can be more easily seen by the following simple argument. For a universe with a homogeneous and isotropic distribution of matter, the gravitational field must be vanishing at every space position since otherwise the direction of the gravitational field ${\bf g}$ would specify a preferred direction at that position. If Newton's law of gravity is written in the form of 
\begin{eqnarray}
        \nabla\cdot{\bf g}=-4\pi G\rho \;, \label{new}
\end{eqnarray}
where $\rho$ is the mass density of matter and $G$ is the gravitational constant, one immediately sees that it is violated since the left-hand side of the equation is zero but the right-hand side is nonzero. 

The same problem is encountered in astrophysics when people consider linear perturbations to a static and uniform system. For a gas of uniform density, pressure, and zero velocity, to the zeroth-order (i.e., for the unperturbed state) the equation of Newton's gravity is violated. Astrophysicists steer by this problem by simply ignoring the unperturbed equation for Newton's gravity and applying equation (\ref{new}) only to the perturbed density and the perturbed gravitational field \cite{bin87,jea29}. This treatment is usually called the Jeans swindle. 

The trouble in application of Newton's law to the universe as a whole, or to a globally uniform system, is essentially caused by the fact that the entire Newton's theory was based on the concept of absolute spacetime. In a universe that is homogeneous and isotropic, at every space position an object must suffer a zero external force so is an inertial frame. According to Newton's first law, any two inertial frames must be at rest relative to each other, or move to or away from each other with a constant velocity. However, because of the existence of mass in the universe, any two inertial frames at a distance should accelerate to each other according to Newton's law of gravity, unless a positive cosmological constant is introduced. This paradox constitutes a fundamental and conceptual problem for describing the dynamics of the universe with Newton's theory.\footnote{Although the dynamic equations for the evolution of and structure formation in a homogeneous and isotropic universe can formally be derived with Newton's law of gravity and motion, it does not deny the fundamental and conceptual problems of Newton's theory in application to the universe as a whole \cite{pee93,wei08}. For example, due to the fact that in Newton's theory light travels at an infinite speed and gravity propagates instantaneously, the formulation of the Newtonian equations applies only to a region smaller than the cosmic horizon. Extension to scales comparable to and larger than the cosmic horizon must include relativistic corrections \cite{gre12,fle12}. In fact, once the distance is extended to cosmological scales, the definition for distance in the Newtonian equations becomes ambiguous since there are multiple and distinct definitions of distance in an expanding universe \cite{wei72}.}

The difficulty of Newton's law of gravity in application to cosmology completely disappeared after Einstein invented the theory of general relativity \cite{ein15a,ein15b} and Hubble discovered that the universe is expanding \cite{hub29}. In fact, general relativity is the first and the unique widely accepted physical theory that can consistently describe the dynamics of the global universe. In general relativity, the usual concept of gravitational field is replaced by that of spacetime curvature: the presence of mass makes spacetime curved and hence affects the motion of objects in the spacetime. The equivalence between a local gravitational field and the acceleration of a point mass makes it possible to choose a frame at any position in a spacetime (the local inertial frame) in which the gravitational field vanishes. Einstein's field equation can describe the dynamics of a homogeneous and isotropic universe without any problem, although it reduces to Newton's equation in the limit of weak fields, slow velocity, and small scales \cite{haw75,wal84}. So, the Neumann-Seeliger paradox does not exist in general relativity. The Jeans swindle can also be properly justified in the framework of an expanding universe, at least in principle \cite{fal13}.

When Maxwell's equations of electromagnetic fields are applied to the universe as a whole, a trouble similar to the Neumann-Seeliger paradox also arises. If we replace the ${\bf g}$ on the left-hand side of equation (\ref{new}) by the electric field ${\bf E}$ and the $-G\rho$ on the right-hand side by the electric charge density $\rho_e$, equation (\ref{new}) becomes the Gauss's law of electromagnetism, i.e., the first equation in the complete set of Maxwell's equations
\begin{eqnarray}
        \nabla\cdot{\bf E}=4\pi \rho_e \;. \label{cou}
\end{eqnarray}
For a homogeneous and isotropic universe filled with a uniformly distributed charge, we have ${\bf E}=0$ everywhere since otherwise ${\bf E}$ would provide a preferred direction. Then Gauss's law is violated, since the right-hand side of equation (\ref{cou}) is nonzero. This indicates an inconsistency of Maxwell's equations with a homogeneous and isotropic universe containing a net charge. One might object to the ratiocination by the arguments listed below, which we will show not to hold.

1. The Gauss's law in equation (\ref{cou}) is an equation in a flat spacetime, but the universe is a curved spacetime. That is true. However, in Sec.~\ref{fail} we will show that in the framework of general relativity the above ratiocination also holds and the inconsistency still exists. In fact, we will see that equation (\ref{cou}) remains valid in the standard Maxwell theory extended to a curved spacetime, if the $\nabla$ is interpreted as a covariant derivative operator on a spatial slice of the curved spacetime.

2. A universe must have a zero total charge then the inconsistency does not exist. However, there is no known first principle to force the total charge in a universe to be exactly zero. At least there is yet no such a principle that we know. In fact, a nonzero total charge can arise from several possibilities, including difference in the magnitude of electron and proton charges, nonvanishing charge of neutrons and neutrinos, and asymmetry of matter and anti-matter. Observational and experimental tests of the charge difference and asymmetry are being persistently carried on by researchers, although positive results have not been obtained yet \cite{sen96,cap05,bre11,amo14}. 

3. Maxwell's equations determine that the total charge in a universe must be exactly zero. However, there is no reason to believe that Maxwell's equations must be correct on cosmological scales. Although gauge invariance is often taken to argue for a massless electromagnetic field, in quantum field theory it is well known that a gauge field can acquire a mass through spontaneous symmetry breaking and the Higgs mechanism \cite{pes95,ryd96}. So an electromagnetic field equation with a photon mass term is possible. Observational and experimental limit on the photon mass are being actively pursued by many researchers (\cite{tu05,gol10}, and references therein).

4. The real universe is not precisely homogeneous and isotropic hence the above argument does not apply. If one considers linear perturbations to a homogeneous and isotropic universe with a uniform distribution of charge, similar to the case of linear perturbations to a static and uniform astrophysical system studied by Jeans one may get the correct first-order perturbed Maxwell equation by the Jeans swindle. However, the zeroth-order Maxwell equation remains unsolved. In addition, as will be shown latter in this paper, in a closed universe with a nonzero charge the Maxwell equation always fails regardless of the symmetry of the spacetime and the distribution of charge.

The inconsistency problem of Maxwell's equations in application to a homogeneous and isotropic universe was already encountered by \citet{lyt59} in 1959, when they explored the physical consequences of a general excess of charge. They proposed that the observed expansion of the universe is caused by the repulsive force arising from a uniformly distributed net charge due to a tiny difference in the magnitude of electron and proton charges. They found that Maxwell's equations must be modified to be consistent with the model of a steady de~Sitter universe, and a steady creation of charge is required (see the comment by \citet{hoy60} and the response by \citet{lyt60}). However, as \citet{bar74} correctly pointed out, in a homogeneous and isotropic universe the electric field must be vanishing as forced by the symmetry of the spacetime, hence there cannot exist a repulsive electromagnetic force. More recent discussions on a charged universe and notes on the inconsistency with Maxwell's equations can be found in Refs. \cite{ori85,bri01,mas02,dol07}, and references therein.

In this paper, the trouble and paradox arising from application of Maxwell's equations to a homogeneous and isotropic universe with a net charge are presented and proved with a rigor analysis in the frame work of general relativity. They are also generalized to a more general case: in a closed universe or spacetime (i.e., the spatial section of the spacetime is a three-dimensional compact manifold) with a nonzero net charge, Maxwell's equations always fail without any requirement on the symmetry of spacetime and of the charge distribution. Then, we investigate how to address the troubles and paradoxes by considering modifications to Maxwell's equations. First, we show that an electromagnetic field equation with a photon mass term (i.e., the Proca equation) can address the paradoxes. Although spontaneous symmetry breaking and the Higgs mechanism can naturally generate a photon mass term in the electromagnetic field equation, we argue that it is not the ultimate solution since the photon mass disappears when the symmetry is restored. Maxwell's equations with a native photon mass term is not favored because they introduce a new and unusually small dimensional parameter (a photon mass).

Then, we consider to modify Maxwell's equations in a more fundamental way by introducing to the field equation a term which couples the electromagnetic field potential vector to the Ricci curvature tensor of the spacetime. This is more attractive than introducing a photon mass term, since it does not introduce any new dimensional parameter. In a Ricci-flat spacetime, including the flat spacetime as a special case, the curvature-coupled term disappears and the electromagnetic field equation returns to the Maxwell equation. This means that in a Ricci-flat spacetime gauge invariance is restored. Since a non-empty universe must not be Ricci-flat, the paradoxes can be addressed by the inclusion of a curvature-coupled term in the electromagnetic field equation. If there is a cosmological constant in the universe, the curvature-coupled term in the electromagnetic field equation implies a photon mass determined by the cosmological constant. In this sense we can say that the new electromagnetic field equation implies the Proca equation. 

Besides the above mentioned inconsistency problem of the standard Maxwell equation in application to a universe as a whole, we have a stronger motivation for inclusion of a curvature-coupled term in the electromagnetic field equation. In general relativity, a law of physics in a flat spacetime (except the law of gravity, of course) is usually generalized to a curved spacetime by the ``minimal substitution rule''. That is, in the law of physics, the Minkowski metric tensor is replaced by the general metric tensor in the curved spacetime, and the ordinary derivative operator is replaced by the covariant derivative operator associated with the general metric tensor \cite{wal84}. There is a well-known ambiguity in this rule. If the law of physics contains second-order derivative operators acting on a vector or a tensor, the order of covariant derivative operators matters in a curved spacetime. In a curved spacetime, when a second-order derivative operator acts on a vector or a tensor, exchange of the order of derivative operators causes the presence of the Riemann curvature tensor in the equation. Because of this ambiguity in the ``minimal substitution rule'', generalization of a law of physics in a curved spacetime is determined only up to addition of a term coupled to the spacetime curvature. Hence, it is natural to consider an electromagnetic field equation with a curvature-coupled term, since the field equation contains second-order derivatives of the potential vector.

The ``minimal substitution rule'' is essentially reflection of the principle of equivalence, which states that at any point in a spacetime it is possible to choose a free-fall frame where in a sufficiently small region the law of physics takes the form as in a flat spacetime \cite{wei72,wal84}. The critical point in the statement of the principle of equivalence is that the region in the question must be sufficiently small: it must have a size much smaller than the radius of the spacetime curvature, so that any term in the law of physics inherently coupled to the spacetime curvature can be ignored. Therefore, the presence of a term coupled to the spacetime curvature in the law of physics does not violate the principle of equivalence. Such a term cannot be recovered by the ``minimal substitution rule'', since it does not show up in a flat spacetime. In a region that is not small compared to the curvature radius of the spacetime, the principle of equivalence breaks down and the effect of a curvature-coupled term can be important. For the case of electromagnetic fields, the presence of a curvature-coupled term in the field equation can represent interaction of electromagnetic fields with the spacetime curvature. In fact, when the standard Maxwell equation in a curved spacetime is expressed as a second-order differential equation of the potential vector (i.e., as a wave equation) a curvature-coupled term also appears \cite{wal84}, whose effects have been studied in \cite{tsa01,tsa05}.

In the paper we also discuss some consequences of the modified field equations, including the time delay in propagation of photons with different energy, effects on the statistical mechanics of photons, and constraint on the charge excess in the universe and the difference in the magnitude of electron and proton charges. We also investigate the testability of the new electromagnetic field equation with a curvature-coupled term. We find that, the effect of the curvature-coupled term can be detectable in an astrophysical environment with a high mass density, or in the early universe when radiation drives the dynamics of cosmic expansion. 

Throughout the paper, we use the geometrized and Planck units with $G=c=\hbar=1$, where $c$ is the speed of light, and $\hbar$ is the reduced Planck constant. However, in a few places the units are restored to get the magnitude of physical quantities.

\section{Failure of Maxwell's equations on cosmological scales}
\label{fail}

As a physical law except the gravitational field equation in a flat spacetime is transplanted to a general curved spacetime, one usually adopts the ``minimal substitution rule'': simply replacing the Minkowski metric $\eta_{ab}$ appearing in the law by the general metric $g_{ab}$, and the derivative operator $\partial_a$ associated with $\eta_{ab}$ by the derivative operator $\nabla_a$ associated with $g_{ab}$. With this minimal substitution approach, Maxwell's equations in a curved spacetime take the form (see, eg., \cite{wal84})
\begin{eqnarray}
        \nabla_aF^{ab}=-4\pi J^b \;, \label{meq0a}
\end{eqnarray}
and
\begin{eqnarray}
        \nabla_{[a}F_{bc]}=0 \;. \label{meq0b}
\end{eqnarray}
Here $F_{ab}$ is the anti-symmetric tensor of the electromagnetic field, $J^a$ is the charge current density 4-vector, and square brackets in the index of a tensor denote anti-symmetrization of the tensor.

In this section, we show that {\em the inhomogeneous Maxwell equation (\ref{meq0a}) fails when it is applied to a homogeneous and isotropic universe with a uniformly distributed net charge}. We call it {\em Type I Paradox}. In addition, {\em if the spatial section of the universe is compact (i.e., the universe is closed), the inhomogeneous Maxwell equation (\ref{meq0a}) always fails if the universe has a nonzero net charge, without any requirement on the symmetry properties of the spacetime and of the charge distribution.} We call it {\em Type II Paradox}.

For a universe with a homogeneous and isotropic distribution of matter, which is considered to be consistent in high precision with all modern observations on our universe, we expect that the electromagnetic field must vanish everywhere. Otherwise, the electric field or the magnetic field will provide a preferred spatial direction at a position in the universe, which conflicts with the assumption that the universe is homogeneous and isotropic (i.e., the so-called {\em cosmological principle} \cite{wei72}). Then we must have $F_{ab}=0$, and by equation (\ref{meq0a}), $J^a=0$ everywhere in the universe. This immediately implies that the inhomogeneous Maxwell equation (\ref{meq0a}) cannot describe a homogeneous and isotropic universe with a uniformly distributed net charge.

The above statement can be more clearly seen if the Maxwell equation (\ref{meq0a}) is converted to the form of Gauss's law. To do so, consider an observer comoving with the expansion of the universe with a 4-velocity $u^a=(\partial/\partial t)^a$, where $t$ is the cosmic time. The electric field and the magnetic field measured by the observer are related to the anti-symmetric tensor $F_{ab}$ by
\begin{eqnarray}
        E_a=F_{ab}u^b \;, \label{Ea}
\end{eqnarray}
and
\begin{eqnarray}
        B_a=-\frac{1}{2}\epsilon_{abcd}u^bF^{cd} \;, \label{Ba}
\end{eqnarray}
respectively, where $\epsilon_{abcd}$ is the totally anti-symmetric tensor of the positively oriented volume element associated with the metric $g_{ab}$. Note, $E_au^a=B_au^a=0$. 

On a spacelike hypersurface $\Sigma_t$ orthogonal to $u^a$ (defined by $t=\mbox{const}$, such a hypersurface always exists for a homogeneous and isotropic universe), we can define a spatial metric on $\Sigma_t$ induced from $g_{ab}$ by $h_{ab}=g_{ab}+u_au_b$, and a spatial derivative operator $D_a$ associated with it. Then, from equation (\ref{meq0a}) we can derive that
\begin{eqnarray}
        D_aE^a=4\pi\rho_e \;, \label{pois}
\end{eqnarray}
where $\rho_e\equiv -u_aJ^a$ is the charge density measured by the observer. For a universe with a uniformly distributed charge, $\rho_e$ is a function of the cosmic time $t$. But, as discussed above, for a homogeneous and isotropic universe we must have $E_a=0$ everywhere. Then equation (\ref{pois}) is violated, if $\rho_e\neq 0$.

In fact, equation (\ref{pois}) holds in any spacetime if $E^a$ is interpreted as the electric field measured by an observer with a 4-velocity $u^a$, $\rho_e$ as the charge density measured by the observer, and $D_a$ as the derivative operator associated with $h_{ab}=g_{ab}+u_au_b$. In this sense, the classical equation (\ref{cou}) remains valid in general relativity, if it is hypothesized that the Maxwell equation (\ref{meq0a}) is correct.

\begin{figure}
\vspace{3pt}
\begin{center}\includegraphics[angle=0,scale=0.8]{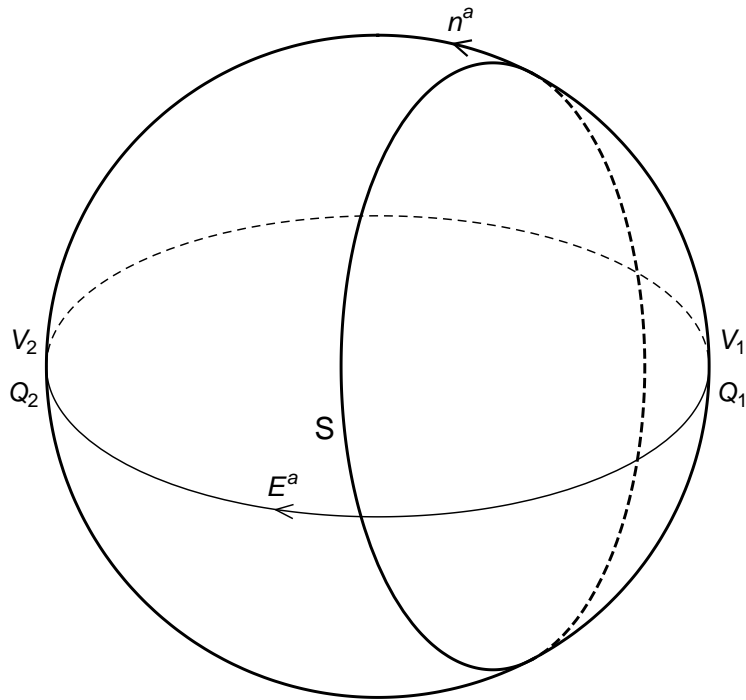}\end{center}
\caption{A compact spacelike hypersurface $\Sigma_t$ in a spacetime $(M,g_{ab})$ is divided into two parts of volume $V_1$ and $V_2$ by a two-dimensional surface $S$. The surface $S$ can be considered as the boundary of $V_1$ with a normal vector $n^a$. It can also be considered as the boundary of $V_2$ with a normal vector $-n^a$. The electric field vector on the surface $S$ is denoted as $E^a$. The total charge contained inside $V_1$ is $Q_1$. The total charge contained inside $V_2$ is $Q_2$. Application of the Gauss's law in equation (\ref{gaus}) to the volume $V_1$ and $V_2$ leads to the conclusion that $Q_1+Q_2=0$.
}
\label{com}
\end{figure}

Equation (\ref{pois}) can be integrated over a 3-volume $V$ enclosed by a two-dimensional surface $S$ on $\Sigma_t$. Then, by Stokes's theorem, we get the Gauss's law for electromagnetism
\begin{eqnarray}
        \frac{1}{4\pi}\int E_an^a dS=Q \;, \label{gaus}
\end{eqnarray}
where $Q\equiv \int_V \rho_e dV$ is the total electric charge contained in the volume $V$. If we apply the Gauss's law in equation (\ref{gaus}) to a volume in a homogeneous and isotropic universe, we get $Q=0$. Hence, {\em Maxwell's equations imply that inside any finite volume of a homogeneous and isotropic universe the total charge must be zero}, conflicting with the assumption that the universe has a nonzero and uniformly distributed net charge. 

With the Gauss's law in equation (\ref{gaus}), a stronger conclusion can be obtained if the hypersurface $\Sigma_t$ is compact (i.e., if the universe is closed). For a compact space $\Sigma_t$, we can use a compact surface $S$ to divide it into two parts: $V_1$ and $V_2$, with $V=V_1+V_2$. Both $V_1$ and $V_2$ are three-dimensional compact manifolds with a boundary surface $S$. We can consider $V_1$ as the interior of $S$, and $V_2$ as the exterior. Alternatively, we can also consider $V_2$ as the interior and $V_1$ as the exterior (Fig.~\ref{com}).

If the Gauss's law in equation (\ref{gaus}) is applied to the space $V_1$ bounded by the surface $S$ with an outward normal $n^a$, as sketched in Fig.~\ref{com}, we get the total charge inside $V_1$: $Q_1=(1/4\pi)\int_S E_an^a dS$. If the Gauss's law is applied to the space $V_2$ bounded by the surface $S$ (then the $n^a$ is an inward normal), we get the total charge inside $V_2$: $Q_2=-(1/4\pi)\int_S E_an^a dS=-Q_1$. Thus, {\em Maxwell's equations imply that the total charge in a closed universe must always be zero}. This statement does not rely on the assumption of homogeneity and isotropy of the universe and uniform distribution of the charge. In other words, Maxwell's equations cannot describe a closed universe with a nonzero net charge without any requirement on spacetime symmetry and charge distribution.

\section{Modification of Maxwell's equations}
\label{em_eq}

In Sec.~\ref{fail} we have shown that Maxwell's equations cannot describe a universe with a nonzero total charge, which indicates that Maxwell's equations have a fundamental flaw when they are applied to the universe as a whole. In this section and the rest part of the paper, we investigate how to modify Maxwell's equations so that the inconsistency problem can be resolved.

If we introduce an electromagnetic potential 4-vector $A^a$ and define the electromagnetic field tensor $F_{ab}$ by
\begin{eqnarray}
        F_{ab}=\nabla_a A_b-\nabla_b A_a \;, \label{F_A}
\end{eqnarray}
the homogeneous Maxwell equation (\ref{meq0b}) is automatically satisfied and hence trivial. Since only the inhomogeneous Maxwell equation (\ref{meq0a}) leads to conflicting results when applied to a universe with a nonzero total charge, we need only modify equation (\ref{meq0a}). Since $F_{ab}=0$ in a homogeneous and isotropic universe, inclusion of additional terms proportional to $F_{ab}$ in the field equation does not help to address the problem. Then, the only possibility left is to add a term proportional to $A^a$ to the field equation, if we want to keep the electromagnetic field equation linear in $A^a$.

The simplest way to modify the Maxwell equation (\ref{meq0a}) so that it can describe a universe with a nonzero total charge is to add a term $-\xi A^a$ to the left-hand side, where $\xi$ is constant. Then, equation (\ref{meq0a}) becomes
\begin{eqnarray}
        \nabla_aF^{ab} -\xi A^b = -4\pi J^b \;. \label{meq1}
\end{eqnarray}
When $\xi$ is positive, this equation is equivalent to the Proca equation \cite{ryd96,tu05,gol10} and $m_A\equiv\xi^{1/2}$ can be interpreted as the rest mass of photons. Although $\xi$ can just be a fundamental constant in nature, an effective $\xi A^a$ term can naturally arise from spontaneous symmetry breaking by the Higgs mechanism if the electromagnetic field is coupled to a complex scalar field \cite{pes95,ryd96}.

With the field equation (\ref{meq1}), the Type I Paradox described in Sec.~\ref{fail} can be addressed. For instance, for an electric charge uniformly distributed in a homogeneous and isotropic universe, the solution to equation (\ref{meq1}) is
\begin{eqnarray}
        A^a=\frac{4\pi}{\xi}\rho_eu^a \;. \label{A_sol1}
\end{eqnarray}
Here $u^a=(\partial/\partial t)^a$ is the 4-velocity of an observer comoving with the expansion of the universe, and $\rho_e=\rho_e(t)$ is the charge density measured by him. The electric current density 4-vector is $J^a=\rho_eu^a$.

By equation (\ref{A_sol1}) we have
\begin{eqnarray}
        \nabla^bA^a=\frac{4\pi}{\xi}\left(\dot{\rho_e}u^au^b+\rho_e\nabla^bu^a\right) \;, 
\end{eqnarray}
where $\dot{\rho_e}\equiv d\rho_e/dt$. For a homogeneous and isotropic universe, $\nabla^bu^a$ is just the extrinsic curvature tensor of the hypersurface defined by $t=\mbox{const}$. Therefore we have $\nabla^bA^a=\nabla^aA^b$ and $F_{ab}=0$, which guarantees that the $A^a$ is consistent with the cosmological principle and solves the electromagnetic field equation (\ref{meq1}). 

Contraction of any 4-velocity $u^a$ with equation (\ref{meq1}) leads to
\begin{eqnarray}
        D_aE^a+\xi\Phi=4\pi\rho_e \;, \label{gaus2}
\end{eqnarray}
where $\Phi=-u_aA^a$ is the scalar potential of electromagnetic fields. Integrating equation (\ref{gaus2}) over the volume $V_1$ and $V_2$ in a compact space (Fig.~\ref{com}) respectively, we get
\begin{eqnarray}
        Q_1+Q_2=\frac{\xi}{4\pi}\int_{V_1+V_2}\Phi dV \;,
\end{eqnarray}
which is not necessarily zero. Hence the Type II Paradox disappears too.

The electromagnetic field equation (\ref{meq1}) can be derived from the action
\begin{eqnarray}
        S_\emm = \int L_\emm \sqrt{-g}\,{\bf e} \;, \label{S_em}
\end{eqnarray}
where the Lagrangian density $L_\emm$ is defined by
\begin{eqnarray}
        L_\emm \equiv -\frac{1}{4}F_{ab}F^{ab}-\frac{\xi}{2} A_aA^a+4\pi A_aJ^a \;.
        \label{L_em}
\end{eqnarray}
Here ${\bf e}$ is a fixed volume element, and $g$ is the determinant of the spacetime metric in the coordinate system compatible with ${\bf e}$. Variation of $S_\emm$ with respect to $A_a$ leads to the field equation (\ref{meq1}). 

Variation of $S_\emm(J^a=0)$ with respect to $g^{ab}$ gives rise to the stress-energy tensor of electromagnetic fields \cite{wal84}
\begin{eqnarray}
	T_{\emm,ab}\equiv-\frac{1}{2\pi\sqrt{-g}}\frac{\delta S_\emm}{\delta g^{ab}}(J^a=0) \;.
        \label{T_em_ab}
\end{eqnarray}
The derived stress-energy tensor of electromagnetic fields described by the Lagrangian density in equation (\ref{L_em}) is
\begin{eqnarray}
	T_{\emm,ab} = ~^{(0)}T_{\emm,ab}+\!~^{(1)}T_{\emm,ab} \;, \label{TEMab}
\end{eqnarray}
where
\begin{eqnarray}
	~^{(0)}T_{\emm,ab}=\frac{1}{4\pi}\left(F_{ac}F_b^{\;\,c}-\frac{1}{4}g_{ab}F_{cd}F^{cd}\right) \label{0TEMab}
\end{eqnarray}
is just the stress-energy tensor of electromagnetic fields in the ordinary Maxwell's theory, and
\begin{eqnarray}
	~^{(1)}T_{\emm,ab} = \frac{\xi}{4\pi}\left(A_a A_b-\frac{1}{2}g_{ab}A_c A^c\right) \label{1TEMab}
\end{eqnarray}
is the additional stress-energy tensor associated with the $\xi A_aA^a$ term in the Lagrangian.

The trace of the electromagnetic stress-energy tensor defined by equations (\ref{TEMab})--(\ref{1TEMab}) is $T_\emm =-\xi A_aA^a/4\pi$, which is nonzero unless $A^a$ is null. The divergence of the electromagnetic stress-energy tensor is calculated to be
\begin{eqnarray}
	\nabla_aT_\emm^{ab}=-J_aF^{ba} + \frac{\xi}{4\pi}A^b\nabla_aA^a \;, \label{dTab0}
\end{eqnarray}
where the field equation (\ref{meq1}) has been applied.

By the identity $\nabla_b\nabla_aF^{ab}=0$, the field equation (\ref{meq1}) implies that
\begin{eqnarray}
        \nabla_aJ^a=\frac{\xi}{4\pi}\nabla_aA^a \;. \label{dJa0}
\end{eqnarray}
When $\xi\neq 0$, the equation of charge conservation, $\nabla_aJ^a=0$, is maintained if and only if
\begin{eqnarray}
        \nabla_aA^a=0 \;, \label{gauge1}
\end{eqnarray}
which is just the Lorentz gauge condition. Substituting equation (\ref{gauge1}) into equation (\ref{dTab0}), we get the usual equation for the Lorentz force
\begin{eqnarray}
	\nabla_aT_\emm^{ab}=-J_aF^{ba} \;. \label{L_force}
\end{eqnarray}

Equation (\ref{L_force}) indicates that when charge is conserved, the $\xi$-term in the stress-energy tensor, i.e., the $\!\!~^{(1)}T_{\emm,ab}$ given by equation (\ref{1TEMab}), has no effect on the dynamics of charged particles. However, the term $\!\!~^{(1)}T_{\emm,ab}$ affects the spacetime curvature through Einstein's field equations.

As an example showing that a photon mass term can arise from spontaneous symmetry breaking and the Higgs mechanism, let us consider a model with a Lagrangian density
\begin{eqnarray}
        L=-\frac{1}{4}F_{ab}F^{ab}-{\cal D}_a\phi\, ({\cal D}^a\phi)^*-V(\phi) \;. \label{L_phi}
\end{eqnarray}
Here, $\phi$ is a complex scalar field, and the asterisk denotes complex conjugate. The gauge covariant derivative operator ${\cal D}_a \equiv \nabla_a+ieA_a$, where $e$ is the coupling parameter defining the interaction of the complex scalar field and the electromagnetic field. The potential function $V(\phi)$ is taken to be of the form
\begin{eqnarray}
        V(\phi)=\frac{\mu^4}{2\lambda}-\mu^2\phi^*\phi+\frac{\lambda}{2}(\phi^*\phi)^2 \;, \label{V_phi}
\end{eqnarray}
where $\mu$ and $\lambda$ are positive and real-valued parameters. The constant term $\mu^4/2\lambda$ in $V(\phi)$ is introduced to ensure that $V(\phi)=0$ at its minimum.

The model described by equations (\ref{L_phi}) and (\ref{V_phi}) is the curved spacetime version of the model in the chapter 20.1 of Ref.~\cite{pes95}. The Lagrangian is obviously invariant under the gauge transformation defined by
\begin{eqnarray}
        A_a\rightarrow A_a+\nabla_a\Gamma(x) \;, \label{g_tran1}
\end{eqnarray}
and
\begin{eqnarray}
        \phi\rightarrow e^{-ie\Gamma(x)}\phi \;, \hspace{1cm} \phi^*\rightarrow e^{ie\Gamma(x)}\phi^* \;, \label{g_tran2}
\end{eqnarray}
where $\Gamma(x)$ is any real-valued function.

Since $\mu^2>0$, the potential $V(\phi)$ acquires a minimum at $|\phi|=\phi_0\equiv\mu/\lambda^{1/2}$. Let us expand the Lagrangian about the vacuum state $\phi=\phi_0$, and write $\phi=\phi_0+[\phi_1(x)+i\phi_2(x)]/\sqrt{2}$, where $\phi_1(x)$ and $\phi_2(x)$ are real-valued functions. The resulting Lagrangian is
\begin{eqnarray}
        L=-\frac{1}{4}F_{ab}F^{ab}-e^2\phi_0^2A_aA^a-\frac{1}{2}\nabla_a\phi_1\nabla^a\phi_1-\mu^2\phi_1^2 -\frac{1}{2}\nabla_a\phi_2\nabla^a\phi_2-\sqrt{2}e\phi_0A^a\nabla_a\phi_2+... \;, \label{L_phi2}
\end{eqnarray}
where ``...'' stands for terms cubic and quartic in $A_a$, $\phi_1$, and $\phi_2$. 

For any given $\phi$, we can always take the advantage of the freedom of gauge transformation to make $\phi$ real. Hence, in this {\em unitary gauge} \cite{pes95,ryd96}, we have $\phi_2=0$. That is, the massless Goldstone bosons go away by a gauge transformation. Then we get a Lagrangian density which contains only {\em physical} particles
\begin{eqnarray}
        L=-\frac{1}{4}F_{ab}F^{ab}-e^2\phi_0^2A_aA^a-\frac{1}{2}\nabla_a\phi_1\nabla^a\phi_1-\mu^2\phi_1^2 +... \;, \label{L_phi3}
\end{eqnarray}
where ``...'' stands for terms cubic and quartic in $A_a$ and $\phi_1$. Clearly, the Lagrangian in equation (\ref{L_phi3}) describes a massive and real scalar field $\phi_1$ interacting with a massive electromagnetic field $A^a$. The mass of the scalar field $\phi_1$ is $m=\sqrt{2}\mu$. The mass of the vector field $A^a$ (i.e., the mass of photons) is
\begin{eqnarray}
        m_A=\sqrt{2}\,e\phi_0=\sqrt{2}\,e\mu/\lambda^{1/2} \;.
\end{eqnarray}

Although spontaneous symmetry breaking and the Higgs mechanism naturally give rise to an electromagnetic field equation in the form of equation (\ref{meq1}), they cannot be considered as the ultimate solution to the inconsistency problem in application of Maxwell's equations to a universe with a net charge. When the symmetry is restored in high energy states of $\phi$, the electromagnetic field becomes massless again and the inconsistency problem reappears. In the next section we will propose another type of modified electromagnetic field equations which, in our opinions, are better than the field equation (\ref{meq1}) and can be considered as a candidate for the ultimate solution for the inconsistency of Maxwell's equations with a universe containing a net charge.

\section{Effective $\xi$-Term from Spacetime Curvature Coupling}
\label{curv}

Here, we propose a new type of modified electromagnetic field equations, where the electromagnetic potential vector $A^a$ is coupled to the Ricci tensor $R_{ab}$ of the background spacetime. We will show that the new field equations are better than the Proca equation and can also solve the paradoxes discussed in Sec.~\ref{fail}.

Let us consider a Lagrangian density of electromagnetic fields with a term coupled to the Ricci tensor of the spacetime
\begin{eqnarray}
        L_\emm=-\frac{1}{4}F_{ab}F^{ab}-\frac{\kappa}{2} R_{ab}A^aA^b+4\pi A_aJ^a \;, \label{L_em2}
\end{eqnarray}
where $\kappa$ is a dimensionless constant of order unity and can be both positive and negative. The difference between this Lagrangian density and that in equation (\ref{L_em}) is that the $(\xi/2) A_aA^a$ term in equation (\ref{L_em}) is replaced by the $(\kappa/2) R_{ab}A^aA^b$ term in equation (\ref{L_em2}). 

Variation with respect to $A_a$ of the action with a Lagrangian density defined by equation (\ref{L_em2}) leads to the electromagnetic field equation
\begin{eqnarray}
        \nabla_a F^{ab}-\kappa R^b_{\;\,a}A^a=-4\pi J^b \;, \label{meq2}
\end{eqnarray}
which contains a term with the potential vector $A^a$ coupled to the Ricci tensor $R_{ab}$.

Let us check that equation (\ref{meq2}) can solve the Type I Paradox described in Sec.~\ref{fail}. For a net charge uniformly distributed in a homogeneous and isotropic universe, the cosmological principle requires that $F_{ab}=0$, $A^a=\Phi(t)u^a$, and $J^a=\rho_e(t)u^a$, where $u^a=(\partial/\partial t)^a$ is the 4-velocity of an observer comoving with the expansion of the universe. By the discussions in Sec.~\ref{em_eq}, the potential vector $A^a=\Phi(t)u^a$ automatically implies $F_{ab}=0$. For a homogeneous and isotropic universe, the Ricci curvature tensor $R_{ab}$ must have a form of $R_{ab}=\alpha(t)u_au_b+\beta(t)h_{ab}$, where $h_{ab}\equiv g_{ab}+u_au_b$, $\alpha(t)$ and $\beta(t)$ are scalar functions of the cosmic time $t$. Therefore, we have $R^b_{\;\,a}A^a=-\alpha\Phi u^b$, since $u_au^a=-1$. Equation (\ref{meq2}) then has the following solution: $\Phi=-(4\pi/\alpha\kappa)\rho_e$, i.e., $A^a=-(4\pi/\alpha\kappa)J^a$.

Similar to the discussion in Sec.~\ref{em_eq}, with the electromagnetic field equation (\ref{meq2}), the Type II Paradox is also addressed. Contraction of any 4-velocity $u^a$ with equation (\ref{meq2}) leads to
\begin{eqnarray}
        D_aE^a-\kappa R_{ab}u^aA^b=4\pi\rho_e \;. \label{gaus2a}
\end{eqnarray}
Summation of the integrals over the volumes $V_1$ and $V_2$ in a compact space (Fig.~\ref{com}) leads to
\begin{eqnarray}
        Q_1+Q_2=-\frac{\kappa}{4\pi}\int_{V_1+V_2} R_{ab}u^aA^b dV \;,
\end{eqnarray}
which in general does not vanish.

The electromagnetic field equation (\ref{meq2}) has the following features: 1. It does not introduce a new and small dimensional parameter, unlike the modified equation (\ref{meq1}). It also does not rely on the existence of another matter field, e.g., a scalar Higgs field. The dimensionless $\kappa$ in equation (\ref{meq2}) is simply a numerical factor of order unity. 2. For a spacetime with $R_{ab}=0$, it becomes the standard Maxwell equation. This means that, unlike in the case of equation (\ref{meq1}), the modification to the Maxwell equation considered here does not affect the electrodynamics in a flat or Ricci-flat spacetime. In particular, in a flat or Ricci-flat spacetime photons remain massless and the gauge invariance is maintained. 3. For a spacetime with $R_{ab}=\Lambda g_{ab}$, where $\Lambda$ is the cosmological constant, the electromagnetic field equation (\ref{meq2}) reduces to equation (\ref{meq1}) with $\xi=\kappa\Lambda$. 

Therefore, we consider the electromagnetic field equation (\ref{meq2}) more attractive and more fundamental than the Proca-type equation. If the universe contains a cosmological constant, by Einstein's field equations we have
\begin{eqnarray}
        R_{ab}=\Lambda g_{ab}+8\pi\left(T_{ab}-\frac{1}{2}Tg_{ab}\right) \;, \label{Rab_Tab}
\end{eqnarray}
where $T_{ab}$ is the stress-energy tensor of matter, and $T=g^{ab}T_{ab}$. Then we get
\begin{eqnarray}
        R_{ab}A^b=\Lambda A_a+8\pi\left(T_{ab}-\frac{1}{2}Tg_{ab}\right)A^b \;, \label{Rab_A^b}
\end{eqnarray}
which indicates that the electromagnetic field equation contains a photon mass term with $\xi=\kappa\Lambda$. In this sense, we can say that the electromagnetic field equation (\ref{meq2}) implies the Proca-type equation with a photon mass determined by the cosmological constant.

Application of $\nabla_b$ to equation (\ref{meq2}) leads to $\nabla_aJ_\eff^a=0$, where
\begin{eqnarray}
        J_\eff^a\equiv J^a-\frac{\kappa}{4\pi}R^a_{\;\;b}A^b \;. \label{J_eff}
\end{eqnarray}
Hence, the conservation of electric charge, $\nabla_aJ^a=0$, implies that the gauge condition
\begin{eqnarray}
        \nabla^a\left(R_{ab}A^b\right) =0 \label{dJ2}
\end{eqnarray}
has to be satisfied.

Derivations of the stress-energy tensor of the electromagnetic field from an action with the Lagrangian density in equation (\ref{L_em2}), and the corresponding divergence, are presented in Appendixes \ref{Tabs} and \ref{dTabs}. Here we only write down the results. The stress-energy tensor can be written as a sum of two terms as in equation (\ref{TEMab}), with $\!~^{(0)}T_{\emm,ab}$ given by equation (\ref{0TEMab}), and 
\begin{eqnarray}
        ~^{(1)}T_{\emm,ab} = \frac{\kappa}{8\pi}\left\{\nabla^c\nabla_c(A_aA_b)-2\nabla^c\nabla_{(a}(A_{b)}A_c)+4A^cR_{c(a}A_{b)}+g_{ab}\left[\nabla_c\nabla_d(A^cA^d)-R_{cd}A^cA^d\right]\right\} \;, \label{Tab2}
\end{eqnarray}
where parenthesis in the index of a tensor denote symmetrization of the tensor. The divergence of the stress-energy tensor is
\begin{eqnarray}
        \nabla^aT_{\emm,ab} = \frac{1}{4\pi}\left[F_{bc}\left(\nabla_aF^{ac}-\kappa R^c_{\;\,a}A^a\right)+\kappa A_b\nabla^c\left(R_{ca}A^a\right)\right] \;. \label{dTab2}
\end{eqnarray}

Substituting the electromagnetic field equation (\ref{meq2}) and its divergence into equation (\ref{dTab2}), we get
\begin{eqnarray}
        \nabla^aT_{\emm,ab} = -F_{ba}J^a+A_b\nabla_a J^a \;. \label{dTab3}
\end{eqnarray}
When the electric charge is conserved, we get the Lorentz force equation (\ref{L_force}). Hence, the electromagnetic force equation is not affected by the presence of a curvature-coupled term in the electromagnetic field equation. Physical meanings of this fact will be discussed in Sec.~\ref{test}.

\section{Motivations for the Curvature-Coupled Electromagnetic Field Equation}
\label{der_ccm}

Equation (\ref{meq0a}) is obtained from generalization of the Maxwell equation in a flat spacetime
\begin{eqnarray}
        \partial_aF^{ab}=-4\pi J^b \label{meq00}
\end{eqnarray}
to a curved spacetime by direct application of the ``minimal substitution rule''. That is, we replace in the equation the ordinary derivative operator $\partial_a$ by the derivative operator $\nabla_a$ associated with the metric tensor $g_{ab}$ in a curved spacetime, and the $F^{ab}$ and $J^a$ by their correspondences in a curved spacetime. In particular, the definition of $F_{ab}$ in a flat spacetime, $F_{ab}=\partial_aA_b-\partial_bA_a$, is replaced by the definition given by equation (\ref{F_A}). 

The ``minimal substitution rule'' provides a simple and convenient machine for converting an equation of physics in a flat spacetime to a curved spacetime. Although it can guarantee that the derived equation in a curved spacetime satisfies the principle of general covariance and returns to the correct equation in a flat spacetime, the ``minimal substitution rule'' cannot guarantee that the equation derived with it must be correct in physics in a curved spacetime. As a simple example, if the true field equation contains a term involving the spacetime curvature, the term is not present in a flat spacetime so cannot be recovered by the ``minimal substitution rule''. A conformally invariant scalar field is just such an example \cite{bir84}. Since the term $R\phi/6$ is missing in a flat spacetime, the equation of a conformally invariant scalar field cannot be derived from its version in a flat spacetime (i.e., the Klein-Gordon equation for a massless scalar field) by the ``minimal substitution rule''.

The ``minimal substitution rule'' has another profound problem. If a field equation in a flat spacetime contains second or higher order derivatives of a vector or a tensor, the corresponding field equation in a curved spacetime derived by the ``minimal substitution rule'' is not unique. Since the order of derivatives of a vector and a tensor matters in a curved spacetime, the ``minimal substitution rule'' can lead to different field equations in a curved spacetime even starting from the same equation in a flat spacetime, if the order of derivatives is arranged in different ways. The Maxwell equation (\ref{meq00}) contains second-order derivatives of the potential vector $A^a$. So, even if the ``minimal substitution rule'' is insisted, different electromagnetic field equations can be obtained in a curved spacetime. 

To see the above point, let us write the Maxwell equation (\ref{meq00}) in terms of $A^a$
\begin{eqnarray}
        \partial_a\partial^aA^b-\partial_a\partial^bA^a=-4\pi J^b \;, 
\end{eqnarray}
which is equivalent to
\begin{eqnarray}
        \partial_a\partial^aA^b-(1+\kappa)\partial_a\partial^bA^a+\kappa\partial^b\partial_aA^a=-4\pi J^b \label{meq00a} 
\end{eqnarray}
for any $\kappa$. Application of the ``minimal substitution rule'' to equation (\ref{meq00a}) gives rise to
\begin{eqnarray}
        \nabla_a(\nabla^aA^b-\nabla^bA^a)-\kappa\left(\nabla_a\nabla^b-\nabla^b\nabla_a\right)A^a=-4\pi J^b \;, 
\end{eqnarray}
which is just equation (\ref{meq2}) by the identity $R^b_{\;\;a}A^a=\left(\nabla_a\nabla^b - \nabla^b\nabla_a\right)A^a$. Therefore, since $\kappa$ can be any number (it can be any function in fact), the ``minimal substitution rule'' can give rise to an infinite number of field equations in a curved spacetime corresponding to the Maxwell equation in a flat spacetime. Although a requirement of gauge invariance picks up equation (\ref{meq0a}), an electromagnetic field equation in the form of (\ref{meq2}) cannot be excluded since gauge invariance is not a physical principle.\footnote{According to Refs.~\cite{gol10} and \cite{sch13}, gauge invariance is not a symmetry of nature. It generates nothing that is observable. While a global symmetry gives rise to a conserved current by Noether's theorem, the local gauge symmetry does not. Gauge invariance only provides a principle for construction of a local theory for describing massless vector particles and a tool for the convenience of computations.} We also remark that there is no observational or experimental evidence for gauge invariance of electromagnetic fields in a curved spacetime with a nonvanishing Ricci curvature.

Although in equation (\ref{meq00a}) it seems a little arbitrary in writing $\partial_a\partial^bA^a=(1+\kappa)\partial_a\partial^bA^a-\kappa\partial^b\partial_aA^a$, there is a natural way to derive a field equation like that in (\ref{meq2}). The value of $\kappa$ can even be determined. To do so, we first show that the Maxwell equation in a flat spacetime can be expressed in terms of a symmetric tensor instead of the anti-symmetric tensor. Let us define a symmetric tensor $H_{ab}$ by
\begin{eqnarray}
        H_{ab}=\partial_aA_b+\partial_bA_a \;. \label{H_A0}
\end{eqnarray}
Then, by the identity
\begin{eqnarray}
        F_{ab}=H_{ab}-2\partial_bA_a \;,
\end{eqnarray}
the Maxwell equation (\ref{meq00}) can be written as
\begin{eqnarray}
        \partial_aH^{ab}-\partial^bH=-4\pi J^b \;, \label{meq_aa5}
\end{eqnarray}
where $H\equiv \eta^{ab}H_{ab}=2\partial_aA^a$.

Now, applying the ``minimal substitution rule'' to equations (\ref{H_A0}) and (\ref{meq_aa5}), we get the corresponding equations in a curved spacetime
\begin{eqnarray}
        H_{ab}=\nabla_aA_b+\nabla_bA_a \;, \label{H_A} 
\end{eqnarray}
and
\begin{eqnarray}
        \nabla_aH^{ab}-\nabla^bH=-4\pi J^b \;, \label{meq_aa6}
\end{eqnarray}
where now $H=g^{ab}H_{ab}=2\nabla_aA^a$. By the identity
\begin{eqnarray}
        \nabla_aH^{ab}-\nabla^bH = \nabla_aF^{ab}+2R^b_{\;\,a}A^a \;,
\end{eqnarray}
equation (\ref{meq_aa6}) is equivalent to
\begin{eqnarray}
        \nabla_aF^{ab}+2R^b_{\;\,a}A^a=-4\pi J^b \;, \label{meq_aa7}
\end{eqnarray}
which is just the equation (\ref{meq2}) with $\kappa=-2$. 

The electromagnetic field equation (\ref{meq_aa6}) can be derived from an action with a Lagrangian density
\begin{eqnarray}
        L_\emm=-\frac{1}{4}\left(H_{ab}H^{ab}-H^2\right)+4\pi A_aJ^a \;, \label{L_em3}
\end{eqnarray}
by variation with respect to $A_a$. It can be checked that the above Lagrangian is identical to that in equation (\ref{L_em2}) with $\kappa=-2$, up to a boundary term which has no contribution to the action integral. So, the derived stress-energy tensor of the electromagnetic field is given by equations (\ref{TEMab}), (\ref{0TEMab}), and (\ref{Tab2}) with $\kappa=-2$.

If we define another symmetric tensor
\begin{eqnarray}
        \Theta_{ab}\equiv H_{ab}-Hg_{ab} \;, \label{Theta_ab}
\end{eqnarray}
the electromagnetic field equation (\ref{meq_aa6}) can be recast in a neater form
\begin{eqnarray}
        \nabla_a\Theta^{ab}=-4\pi J^b \;. \label{meq4t}
\end{eqnarray}
The Lagrangian density in equation (\ref{L_em3}) can be expressed with $\Theta_{ab}$ as
\begin{eqnarray}
        L_\emm=-\frac{1}{4}\left(\Theta_{ab}\Theta^{ab}-\frac{1}{3}\Theta^2\right)+4\pi A_aJ^a \;, \label{L_em3a}
\end{eqnarray}
where $\Theta\equiv g^{ab}\Theta_{ab}=-3H$.

The above arguments support the proposal of equation (\ref{meq2}) as a candidate for the electromagnetic field equation in a curved spacetime. The paradoxes arising from application of Maxwell's equations to a universe as described in Sec.~\ref{fail} demands the inclusion of the curvature-coupled term in the electromagnetic field equation. The discussions in this section also explain why we consider to add a curvature-coupled term $R^b_{\;\;a}A^a$ instead of a simpler $RA^b$ to the field equation.

\section{Some Consequences of the Modified Electrodynamics}
\label{cons}

Both the two types of modified Maxwell's equations discussed in previous sections have an unavoidable consequence: photons can have a nonzero rest mass. Equation (\ref{meq1}) indicates an intrinsic photon mass, if $\xi$ is interpreted as a fundamental constant. Equation (\ref{meq2}) indicates an effective photon mass as a function of the local spacetime curvature. For example, as mentioned in Sec.~\ref{curv}, if there exists a cosmological constant as current observations indicated \cite{dun09,pla15}, the $\kappa R^b_{\;\;a}A^a$ term in equation (\ref{meq2}) gives rise to a photon mass $m_A=(\kappa\Lambda)^{1/2}$, corresponding to $\lambda_A\sim 10^{28}\cm$ (the size of the cosmic horizon). Here $\lambda_A=2\pi\lambdabar_A= 2\pi\hbar/m_A c$ is the Compton wavelength of a massive photon.

At present the most stringent and secure limit on the photon mass comes from the measurement of the magnetic field in the solar wind: $m_A<2\times 10^{-51}\g$, or $\lambdabar_A>2\times 10^{13}\cm$ \cite{ryu07}. A much stronger but very speculative constraint comes from observations of the magnetic field on galactic scales, which is $m_A<10^{-59}\g$, or $\lambdabar_A>3\times 10^{21}\cm$ \cite{ade07}. In this section we investigate some consequences of a nonzero photon mass, or equivalently, of a positive $\xi$. For simplicity, we assume that $m_A=\xi^{1/2}$ is a constant.

\subsection{Propagation time delay of photons with different energy}
\label{prop}

An immediate consequence of a photon mass is the appearance of dispersion in the velocity of photons: bluer photons travel faster, and the velocity approaches $c$ as the frequency of photons approaches infinity. However, attempts in measuring or giving a limit on the photon mass by observing the difference in the arriving time of photons with different frequencies emitted by the same source are disfavored, since for any reasonable value of a photon mass the effect turns out to be extremely small \cite{deb40,gin64,kob68}. 

For example, for a radio source at a cosmological distance $D$, the time delay of radio photons relative to a photon traveling with a speed $c$ is
\begin{eqnarray}
	\Delta t = 1.50{\rm s} \left(\frac{D}{10^{28}{\rm cm}}\right)\left(\frac{\lambda_A}{10^{10}{\rm cm}}\right)^{-2}\left(\frac{f}{10^9{\rm Hz}}\right)^{-2} \;, \label{dt2}
\end{eqnarray}
where $f=\omega/2\pi$ is the frequency of a radio photon, and the cosmological distance $D$ is defined by
\begin{eqnarray}
        D_X(z)=\int_0^{z}\frac{cdz}{H(z)(1+z)^2} \;. \label{Dx}
\end{eqnarray}
Here $z$ is the cosmic redshift, $H\equiv d\ln a/dt$, where $a=a(t)$ is the linear scale of the universe at time $t$.\footnote{The distance defined by equation (\ref{Dx}) does not appear to be identical to any existing distance definition in cosmology, therefore we denote it by $D_X$.}

Taking $\lambda_A\gtrsim 10^{14}\cm$, we get $\Delta t\lesssim 10^{-8}\s$ for a radio source at a cosmological distance. This makes it almost impossible to detect the delay in the arriving time of photons due to a nonzero photon mass by observing the radio emissions of a gamma-ray burst or a quasar. 

\subsection{Statistical mechanics of massive photons}
\label{stat}

There is a profound discontinuity as the mass of photons approaches zero: massive photons have three degrees of freedom, but massless photons have only two. In this subsection we discuss the effect of a photon mass on the statistical mechanics of photons. To be specific, we assume that the background spacetime is a Minkowski spacetime.

Massless photons obey the Bose-Einstein statistics and in the thermal equilibrium state have a spectrum of energy density given by the Planck radiation law
\begin{eqnarray}
        u(\omega)=\frac{g\hbar}{2\pi^2c^3}\frac{\omega^3}{e^{\hbar\omega/kT}-1} \;. \label{pl_law}
\end{eqnarray}
Here $T$ is the temperature, $k$ is the Boltzmann constant, and $g$ is the statistical degeneracy factor. For massless photons we have $g=2$.

Massive photons are described by the statistics of relativistic Bose gases \cite{ara80,moh83}. When $kT\gg m_A c^2$, the rest mass of photons can be ignored and the statistics approaches that of massless photons. The total number of photons are not conserved and the spectrum of energy density is given by equation (\ref{pl_law}). The question is in the value of $g$. Naively, one may expect that $g=3$ since massive photons have a longitudinal component (i.e., a spin component with $s=0$) that massless photons do not have. However, the value of $g$ depends on how the longitudinal photons interact with matter. If they interact with matter sufficiently weak or do not interact with matter at all, we still get $g=2$ and the Stephan-Boltzmann law remains valid \cite{tu05,gol10,sch06}.

Here we demonstrate that when photons have an energy $\hbar\omega\gg m_A c^2$ they interact with matter very weakly by considering the electromagnetic radiation generated by an electric dipole ${\bf p}$ oscillating with a frequency $\omega$. Far from the dipole, the three-dimensional vector potential ${\bf A}$ and the scalar potential $\Phi$ are respectively
\begin{eqnarray}
        {\bf A}(t, {\bf x})= -\frac{i\omega}{r}e^{-i\omega t+ikr}{\bf p} \;, \label{A_sol_dp2}
\end{eqnarray}
and
\begin{eqnarray}
        \Phi(t,{\bf x})= -\frac{ik}{r}e^{-i\omega t+ikr}{\bf p}\cdot{\bf n} \;, \label{Phi_sol_dp}
\end{eqnarray}
where $r$ is the distance from the dipole, ${\bf n}$ is the unit vector along the radial direction, and $k=\sqrt{\omega^2-m_A^2}$ is the magnitude of the three-dimensional wavevector.

The magnetic field and the electric field derived from the ${\bf A}$ and $\Phi$ are respectively
\begin{eqnarray}
        {\bf B}=\frac{\omega k}{r}e^{-ik\omega+ikr}\left({\bf n}\times{\bf p}\right) \;, \label{B_sol_dp}
\end{eqnarray}
and
\begin{eqnarray}
        {\bf E} = \frac{k}{\omega}{\bf B}\times {\bf n}+\frac{m_A^2}{r}e^{-i\omega t+ikr}{\bf p} \;. \label{E_sol_dp}
\end{eqnarray}
In the above expressions we have ignored terms of order $r^{-2}$ and higher, since they do not contribute to the Poynting flux and the radiation power.

The magnetic field ${\bf B}$ has only transverse components (i.e., ${\bf B}\cdot{\bf n}=0$). However, in addition to transverse components, the electric ${\bf E}$ has a longitudinal component ${\bf E}_\parallel$ (i.e., ${\bf E}_\parallel\times{\bf n}=0$) given by
\begin{eqnarray}
        {\bf E}_\parallel=\frac{m_A^2e^{-i\omega t+ikr}}{r}({\bf p}\cdot{\bf n}){\bf n} \;. \label{E_sol_dp2_paral}
\end{eqnarray}

From the expressions of ${\bf B}$ and ${\bf E}$, we can calculate the time-averaged Poynting flux vector. When $\xi=m_A^2\neq 0$, a generalized Poynting flux vector is derived in Appendix \ref{EB} and given by equation (\ref{cJ_mom}) (see also \cite{bas55}). Setting $\alpha=1$ in equation (\ref{cJ_mom}), we get the generalized Poynting flux vector
\begin{eqnarray}
        {\bf S} = \frac{1}{4\pi}\left[\langle\Re({\bf E})\times\Re({\bf B})\rangle+m_A^2\langle\Re({\Phi})\Re({\bf A})\rangle\right] = \frac{\omega^3 k}{8\pi r^2}p^2\sin^2\theta\left(1+\frac{m_A^2}{\omega^2}\cot^2\theta\right)\;{\bf n} \;,
\end{eqnarray}
where $\langle\rangle$ denotes time-average, and $\Re(f)$ denotes the real part of a complex function $f$. As expected, ${\bf S}$ has only a component in the $r$-direction.

The power of the dipole radiation is given by the integration of ${\bf S}$ over a sphere of radius $r$, i.e., 
\begin{eqnarray}
        P =r^2\int {\bf S}\cdot{\bf n}d\Omega =\frac{1}{3}p^2\omega^3 k\left(1+\frac{m_A^2}{2\omega^2}\right) \;. \label{power}
\end{eqnarray}
As expected, the power is independent of the radius $r$ as demanded by the law of energy conservation.

Equation (\ref{power}) indicates that, in the radiation generated by an electric dipole, the power components in each photon spin state ($s=1,-1$, and 0) have a ratio
\begin{eqnarray}
        P_{s=1}:P_{s=-1}:P_{s=0} = 1: 1: \frac{m_A^2}{\omega^2} \;. \label{pr}
\end{eqnarray}
Therefore, when $\omega\gg m_A$, the contribution of longitudinal photons to the total radiation is negligible. For example, if $m_A\lesssim 3.5\times 10^{-52}\g$ (corresponding to $\lambdabar_A\gtrsim 10^{14}\cm$) and $\omega= 10^{12}{\rm Hz}$ (corresponding to the peak frequency of the cosmic microwave background) , we get $m_A/\omega\lesssim 3\times 10^{-16}$ and $m_A^2/\omega^2\lesssim 10^{-31}$. We cannot imagine any kind of electromagnetic radiation generated in lab or astrophysical conditions where longitudinal photons can have a detectable contribution.

The relation in equation (\ref{pr}) indicates that longitudinal and transverse photons are not statistically independent and do not have identical statistical weights. For a given statistical state, the number of longitudinal photons are always suppressed by a factor $m_A^2/2\omega^2$ relative to the number of transverse photons. Because of the very weak interaction of longitudinal photons with matter, approaching to a statistical equilibrium state of longitudinal photons in a cavity is extremely slow, with a time-scale comparable to the age of the universe \cite{bas55,stu57,gol71}. According to \cite{gol71}, the practical impossibility for longitudinal photons to get equally partitioned with transverse photons in energy is due to the very large skin depth of longitudinal photons compared to that of transverse photons (with a ratio $\ge \omega^2/m_A^2$) \cite{ger69,kro71}.

So, when $kT\gg m_A c^2$, we expect that for radiation containing both transverse and longitudinal photons, in the equilibrium state transverse photons have an energy spectrum determined by equation (\ref{pl_law}) with $g=2$, and the spectrum of longitudinal photons should be jointly determined by the spectrum of transverse photons and the power ratio relation given by equation (\ref{pr}). Hence, for $\hbar\omega\gg m_A c^2$, the energy spectrum of longitudinal photons should be
 \begin{eqnarray}
        u_\parallel(\omega) = \frac{m_A^2c^4}{2\hbar^2\omega^2}u_\perp(\omega) = \frac{m_A^2c}{2\pi^2\hbar}\frac{\omega}{e^{\hbar\omega/kT}-1} \;, \label{pl_0}
\end{eqnarray}
where $u_\perp(\omega)$ is the energy spectrum of transverse photons (eq.~\ref{pl_law}).

Finally, we argue that it is quite general that the contribution of $m_A^2$ to the radiation field is $\sim (m_A^2/k^2)\times$ the dominant $1/r$ term as indicated in the dipole radiation case (eqs.~\ref{B_sol_dp} and \ref{E_sol_dp}). In a Minkowski spacetime, from equation (\ref{meq1}) we can derive a generalized version of Amp\`ere's circuital law in the differential form
\begin{eqnarray}
        \nabla\times {\bf B}-\frac{\partial{\bf E}}{\partial t}=4\pi \left({\bf J}-\frac{m_A^2}{4\pi}{\bf A}\right) \;. \label{amp_law}
\end{eqnarray}
Since $\partial{\bf E}/\partial t\sim-i\omega{\bf E}$ and ${\bf J}=0$ outside the source, from equation (\ref{amp_law}) we derive that
\begin{eqnarray}
        {\bf E}\sim\frac{i}{\omega}\left(\nabla\times {\bf B}+m_A^2{\bf A}\right) \;. \label{E_sol}
\end{eqnarray}

Since $|\nabla\times{\bf B}|\sim k |{\bf B}|$ and $|{\bf B}|=|\nabla\times{\bf A}|\sim k |{\bf A}|$, by equation (\ref{E_sol}) we get $|{\bf E}|~(\mbox{from~}{\bf B})\sim (k/\omega)|{\bf B}| \sim (k^2/\omega)|{\bf A}|$, and $|{\bf E}|~(\mbox{from~}m_A^2{\bf A})\sim (m_A^2/\omega)|{\bf A}|$. Therefore we have
\begin{eqnarray}
        \frac{|{\bf E}|~(\mbox{from~}m_A^2{\bf A})}{|{\bf E}|~(\mbox{from~}{\bf B})}\sim \frac{m_A^2}{k^2} \;.
\end{eqnarray}

When $kT\ll m_A c^2$ we should have $k^2\ll m_A^2$ and $\omega^2\approx m_A^2$. Then, by equation (\ref{pr}) we get $P_{s=1}:P_{s=-1}:P_{s=0} = 1: 1:1$. Hence, when $kT\ll m_A c^2$, the statistics of massive photons become identical to that of a non-relativistic Bose gas with a degeneracy factor $g=3$. As $T\rightarrow 0$, quantum effects become important and the Bose-Einstein condensation may occur \cite{sch06}.

\subsection{Constraint on the net charge in the universe and the charge difference between electrons and protons}
\label{charge}

Assume that the universe has a mean net charge density
\begin{eqnarray}
        \rho_e=\chi G^{1/2}\rho_m \;, \label{chi}
\end{eqnarray}
where $\rho_m$ is the mean mass density, and the dimensionless parameter $\chi$ is the ratio of charge to mass in the universe. The corresponding charge current density 4-vector is $J^a=\rho_eu^a$, where $u^a=(\partial/\partial t)^a$ is the 4-velocity vector of a fluid comoving with the expansion of the universe. For a homogeneous and isotropic universe, all quantities are functions of the cosmic time $t$. 

As discussed in Secs.~\ref{fail} and \ref{em_eq}, in a homogeneous and isotropic universe we must have $F_{ab}=0$ everywhere. The solution to the electromagnetic equation (\ref{meq1}) is given by equation (\ref{A_sol1}). Then, by equations (\ref{TEMab})--(\ref{1TEMab}), the stress-energy tensor of the electromagnetic field is
\begin{eqnarray}
        T_\emm^{ab} = \frac{4\pi}{\xi}\rho_e^2\left(u^au^b+\frac{1}{2}g^{ab}\right) \;,
\end{eqnarray}
corresponding to a perfect fluid with mass density $\rho_A$ and pressure $P_A$ defined by
\begin{eqnarray}
        P_A=\rho_A c^2=\frac{2\pi}{\xi}\rho_e^2 \;. \label{rho_emm}
\end{eqnarray}

Substituting equation (\ref{chi}) into equation (\ref{rho_emm}), we get $\rho_A =2\pi Gc^{-2}\chi^2\rho_m^2\lambdabar_A^2$, where $\lambdabar_A=\xi^{-1/2}$. By $\rho_m=\Omega_m\rho_\crit(1+z)^3$, where $z$ is the cosmic redshift, and $\rho_\crit=1.9\times 10^{-29}h^2\,\g\,\cm^{-3}$ is the critical mass density of the universe, we get
\begin{eqnarray}
        \omega_A\equiv\frac{\rho_A}{\rho_\crit}=\frac{3}{4}\chi^2\Omega_m^2\left(\frac{\lambdabar_A}{d_H}\right)^2(1+z)^6 \;. \label{om_A}
\end{eqnarray}
Here the Hubble distance $d_H=c/H_0\approx 10^{28}h^{-1}\cm$, and the Hubble constant $H_0=100h\,\km\,\s^{-1}\Mpc^{-1}$. If the value of $\omega_A$ at the present time (i.e., at $z=0$) is denoted by $\Omega_A$, we find $\omega_A=\Omega_A(1+z)^6$.

By the state equation (\ref{rho_emm}), $\rho_A$ evolves with the cosmic redshift by $\rho_A\propto(1+z)^6$, i.e., $\rho_e\propto(1+z)^3$. Since $\rho_m\propto (1+z)^3$, by equation (\ref{chi}) the parameter $\chi$ remains a constant as the universe evolves.

Modern observations have confirmed that the matter content in the present universe is dominantly composed of about $70\%$ dark energy (with $P=-\rho c^2$), about $30\%$ of cold dark matter and baryonic matter (with $P\ll \rho c^2$), and a very small fraction of radiation (about $10^{-5}$ in mass density with $P=\rho c^2/3$) \cite{dun09,pla15}. Since the state equation $P_A=\rho_A c^2$ differs from all the matter contents just mentioned, comparison of $\rho_A$ with the mass density of various matter contents in the universe will provide a constraint on the value of $\chi$. 

For example, the requirement of $\rho_A<\rho_m$ at $z=0$ leads to $\chi<2(3\Omega_m)^{-1/2}d_H/\lambdabar_A\approx 2.1d_H/\lambdabar_A$. The requirement of $\rho_A<\rho_r$ (the mean mass density of radiation) at $z=0$ leads to $\chi<2(\Omega_r/3)^{1/2}\Omega_m^{-1}d_H/\lambdabar_A\approx 0.01d_H/\lambdabar_A$. For comparison, $\chi\approx 10^{21}$ and $10^{18}$ for an electron and a proton, respectively.

A net charge in a universe can arise from several possibilities, including a difference in the fundamental charge of electrons and protons which together with neutrons form the fundamental blocks of ordinary matter, a nonvanishing charge of neutrons and neutrinos, and asymmetry of matter and anti-matter. Hence, a constraint on the parameter $\chi$ can be converted to a constraint on the charge difference between electrons and protons, the charge of neutrons and neutrinos, and the asymmetry of matter and anti-matter \cite{lyt59,sen96,cap05}. 

In the original paper of \citet{lyt59}, they (mistakenly) claimed that a small difference in the magnitude of electron and proton charges may provide a repulsive force for driving the expansion of the universe. Based on the then available observational data, they concluded that a fractional difference $y\approx 2.2\times 10^{-18}$ would be sufficient to explain the observed expansion. However, as we have already demonstrated, in a homogeneous and isotropic universe $F_{ab}$ must vanish everywhere so there is no electromagnetic force on the cosmological scale. In fact, the stress-energy tensor of the electromagnetic field arising from the net charge causes an attractive force, instead of a repulsive force according to Einstein's field equation. 

The approach of Lyttleton \& Bondi can still be used to constrain the difference in the magnitude of electron and proton charges. An electron has a charge $-|e|$. Assuming that a proton has a charge $(1+y)|e|$, where $|y|\ll 1$. For a universe containing an equal number of electrons and protons, the total net charge density in the universe at any redshift $z$ is given by $\rho_e=y\rho_b|e|/m_p$, where $m_p$ is the proton mass, and $\rho_b$ is the average mass density of baryons in the universe. Comparison with equation (\ref{chi}) yields
\begin{eqnarray}
        \chi=y\chi_p\frac{\Omega_b}{\Omega_m} \;, \label{chi_y}
\end{eqnarray}
where $\chi_p=|e|/G^{1/2}m_p=1.11\times 10^{18}$ is the charge to mass ratio of a proton, and $\Omega_b=\rho_b/\rho_\crit$. Substituting equation (\ref{chi_y}) into equation (\ref{om_A}), we get
\begin{eqnarray}
        \omega_A = \frac{3}{4}y^2\chi_p^2\Omega_b^2\left(\frac{\lambdabar_A}{d_H}\right)^2(1+z)^6 \;. \label{rho_e2}
\end{eqnarray}

The requirement that $\omega_A(z=0)=\Omega_A<\Omega_m$ provides the first constraint on the value of $y$
\begin{eqnarray}
        y < \frac{2\Omega_m^{1/2}}{3^{1/2}\chi_p\Omega_b}\left(\frac{\lambdabar_A}{d_H}\right)^{-1} \approx 1.3\times 10^{-17}\left(\frac{\lambdabar_A}{d_H}\right)^{-1} \;, 
\end{eqnarray}
where we have adopted $\Omega_m=0.3$, $\Omega_b=0.043$, and $h=0.7$ \cite{dun09,pla15}.

A stronger constraint comes from the requirement that $\Omega_A<\Omega_r$. Adopting $\Omega_r=4.2 \times 10^{-5}h^{-2}$ (including the massless neutrinos and the cosmic microwave background \cite{pee93,wei08}), we get
\begin{eqnarray}
        y<\frac{2\Omega_r^{1/2}}{3^{1/2}\chi_p\Omega_b}\left(\frac{\lambdabar_A}{d_H}\right)^{-1}\approx 2.2\times 10^{-19}\left(\frac{\lambdabar_A}{d_H}\right)^{-1} \;. 
\end{eqnarray}

An even stronger constraint on $y$ comes from the condition of $\omega_A<\Omega_r(1+z)^4$ at $z=1100$ (the redshift of cosmic recombination), where we have used the fact that the density of radiation $\rho_r\propto(1+z)^4$. Since $\omega_A\propto(1+z)^6$, we get
\begin{eqnarray}
        y < \frac{2\Omega_r^{1/2}}{3^{1/2}\chi_p\Omega_b}\left(\frac{\lambdabar_A}{d_H}\right)^{-1}(1+z)^{-1} \approx 2.0\times 10^{-22}\left(\frac{\lambdabar_A}{d_H}\right)^{-1} \;. \label{y_con}
\end{eqnarray}
The above constraint on the value of $y$ is consistent with the result obtained in lab experiments if we take $\lambdabar_A\sim d_H\sim 10^{28}\cm$ \cite{bre11}. 

The estimate of the constraint on $y$ presented above is based on the electromagnetic field equation (\ref{meq1}) with a constant $\xi$. If $\xi$ evolves with the cosmic time the above analyses must be accordingly modified. We also note that the above constraints are based simply on comparison of $\rho_A$ with the matter and radiation density in the universe. More complex analyses, e.g., comparison with the temperature fluctuations in the cosmic microwave background, may lead to stronger constraints on the charge asymmetry and the neutrino charge \cite{sen96,cap05}.

\section{On the testability of the field equation (\ref{meq2})}
\label{test}

In Sec.~\ref{curv} we have shown that when there is a cosmological constant $\Lambda$ in the universe, the electromagnetic field equation (\ref{meq2}) contains a photon mass term with $m_A=(\kappa \Lambda)^{1/2}$, corresponding to a Compton wavelength comparable to the Hubble distance. Given the very small value of a possible cosmological constant in the present universe \cite{dun09,pla15}, the effect of a so small photon mass may never be observable, unless the universe has a nonzero net charge (Sec.~\ref{cons}). However, in the epoch of inflation before the Big Bang when the expansion of the universe is presumably driven by a large effective cosmological constant, it can be imagined that a correspondingly large photon mass may affect the spectrum of the Gibbons-Hawking radiation \cite{gib77}. 

Equation (\ref{meq2}) indicates that local curvature of the background spacetime can affect the electrodynamics. So, we can expect that, in a strongly curved spacetime, the effect of the curvature-coupled term may be measurable on scales comparable to the curvature radius of the spacetime. In this section, we estimate the order of magnitude of the curvature-coupled term, and find what kind of environment is favorable for detection of the effect.

By equations (\ref{meq2}) and (\ref{J_eff}), the curvature term can be regarded as a pseudo-charge current density vector
\begin{eqnarray}
        J_{\rm ps}^a=-\frac{\kappa}{4\pi}R^a_{\;\;b}A^b \;. \label{Js}
\end{eqnarray}
It is not a true electric charge current density, but its effect in generation of electromagnetic fields is equivalent to that of an electric charge current density. If an observer measures the electric field outside a charged ball and interprets the result with the standard Gauss's law, he will find that the electric field appears as being generated by a total charge
\begin{eqnarray}
        Q_{\rm tot}=Q+\frac{\kappa}{4\pi}\int_V R_{ab}u^aA^bdV \;. \label{Q_tot}
\end{eqnarray}
Here $Q$ is the electric charge of the ball, $u^a$ is 4-velocity vector of the observer, and $V$ is the volume of the ball.

Similarly, if an observer measures the magnetic field outside a star and interprets the result with the standard Amp\`ere's circuital law, he will find that the magnetic field appears as being generated by a total current density inside the star given by
\begin{eqnarray}
        j^a_{\rm tot}=j^a-\frac{\kappa}{4\pi} h^{ab}R_{bc}A^c \;, \label{j_tot}
\end{eqnarray}
where $j^a$ is the three-dimensional electric current density, and $h^{ab}=g^{ab}+u^au^b$. Therefore, even if the outside of an object is vacuum and Ricci-flat, the nonvanishing curvature-coupled term in the interior of the object can still affect the structure of the electromagnetic field outside the object. If the interior of the object is strongly curved due to the presence of a large mass density, the effect of the pseudo-charge current density arising from the curvature-coupled term may be testable in measurements of the outside electromagnetic field.

To estimate the order of magnitude of the curvature-coupled term, let us ignore the cosmological constant and assume that the matter inside an object is described by a stress-energy tensor $T^{ab}=\rho u^au^b$, where $\rho$ is the mass density of the matter. Then, equation (\ref{Rab_A^b}) becomes
\begin{eqnarray}
        R_{ab}A^b=4\pi\rho\left(A_a-2\Phi u_a\right) \;, \label{Rab_A1}
\end{eqnarray}
where $\Phi=-u_aA^a$ is the scalar electric potential. 

By Einstein's field equations, the curvature radius of a mass system can be estimated by
\begin{eqnarray}
        r_c\sim l_{\rm P}\left(\frac{8\pi\rho}{\rho_{\rm P}}\right)^{-1/2}\sim 10^{13}\cm\left(\frac{\rho}{1\g\,\cm^{-3}}\right)^{-1/2} , \label{r_c}
\end{eqnarray}
where $l_{\rm P}=1.6\times 10^{-33}\cm$ is the Planck length, and $\rho_{\rm P}=5.2\times 10^{93}\g\,\cm^{-3}$ is the Planck mass density. Hence, inside an object, the order of magnitude of $R_{ab}A^b$ can be estimated by $\left|R_{ab}A^b\right|\sim|A^b|/r_c^2$, where the curvature radius $r_c$ is estimated by equation (\ref{r_c}). The first term in equation (\ref{meq2}) can be estimated by $\left|\nabla_aF^{ab}\right|\sim|A^b|/r_A^2$, where $r_A$ is the scale over which the electromagnetic field changes. Hence, we have
\begin{eqnarray}
        \frac{\left|R_{ab}A^b\right|}{\left|\nabla_aF^{ab}\right|}\sim\left(\frac{r_A}{r_c}\right)^2 \;. \label{R_dF_r}
\end{eqnarray}

Equation (\ref{R_dF_r}) indicates that the curvature term in the field equation (\ref{meq2}) will lead to measurable effects if $r_c\lesssim r_A$. Objects clearly satisfying this criterion include neutron stars and the early universe. A typical neutron star has a radius $r\approx 10^6\cm$ and an overall mass density $\rho\approx 5\times 10^{14}\g\,\cm^{-3}$, corresponding to a curvature radius $r_c\approx 10^6\cm$ by equation (\ref{r_c}). Hence, we have $r_c\sim r_A$ for an electromagnetic field with $r_A\sim r$. Therefore, for a neutron star, we expect that the curvature-coupled term in the electromagnetic field equation can produce detectable effects on its inside and outside magnetic fields. In the very early time, the universe has a very high mass density and a very small curvature radius. The curvature-coupled term in the electromagnetic field equation should affect all the electromagnetic processes on scales of the curvature radius of the early universe. 

In the theory of electrodynamics defined by the Lagrangian in equation (\ref{L_em2}), the stress-energy tensor of electromagnetic fields contains an extra term in addition to the usual stress-energy tensor $\!\!~^{(0)}T_{\emm,ab}$. The extra term $\!\!~^{(1)}T_{\emm,ab}$ is defined by equation (\ref{Tab2}). By equations (\ref{dPi}) and (\ref{Js}), the divergence of $\!\!~^{(1)}T_{\emm,ab}$ is
\begin{eqnarray}
        \nabla^a\!\!~^{(1)}T_{\emm,ab} = F_{ba}J^a_{\rm ps}-A_b\nabla_aJ^a_{\rm ps} \;. \label{dT1}
\end{eqnarray}
The $\!\!~^{(1)}T_{\emm,ab}$ only interacts with the pseudo-charge current. Hence, the extra stress-energy tensor $\!\!~^{(1)}T_{\emm,ab}$ represents a type of dark electromagnetic energy and momentum. It does not interact with electric charges and currents. 

In contrast, the stress-energy tensor $\!\!~^{(0)}T_{\emm,ab}$ (eqs.~\ref{0TEMab} and \ref{Tab_EB}) interacts with both the charge current $J^a$ and the pseudo-charge current $J^a_{\rm ps}$, since by equations (\ref{dT0}) and (\ref{meq2}) we have
\begin{eqnarray}
        \nabla^a\!\!~^{(0)}T_{\emm,ab} = -F_{ba}\left(J^a+J^a_{\rm ps}\right) \;. \label{dT00}
\end{eqnarray}
In the total stress-energy tensor of electromagnetic fields, only the $\!\!~^{(0)}T_{\emm,ab}$ part is accessible to experiments based on electromagnetic methods. In a spacetime region that is strongly curved, the conservation of $\!\!~^{(0)}T_{\emm,ab}$ is affected by the presence of a pseudo-charge current, according to equation (\ref{dT00}).

Although $\!\!~^{(1)}T_{\emm,ab}$ does not interact with electric charges, it affects the spacetime geometry through Einstein's field equations. Note, $\!\!~^{(1)}T_{\emm,ab}$ contains terms not involving the spacetime curvature (eq.~\ref{Tab2}). Even in a place where $r_A\ll r_c$ (hence the curvature terms in eq.~\ref{Tab2} are small), the magnitude of $\!\!~^{(1)}T_{\emm,ab}$ can be comparable to that of $\!\!~^{(0)}T_{\emm,ab}$. In the early universe when the gravity of radiation drives the expansion of the universe, the dark electromagnetic energy may play a very important role.

\section{Summary and Conclusions}
\label{sum}

Based on convincing arguments we have verified that Maxwell's equations cause troubles and paradoxes as they are applied to the universe as a whole. Maxwell's equations cannot describe the electrodynamics of a homogeneous and isotropic universe with a uniformly distributed net charge, since in a homogeneous and isotropic universe the electromagnetic field must be vanishing everywhere (Type I Paradox). For a closed universe (i.e., a universe with a compact spatial section), Maxwell's equations always fail if the total charge in the universe is nonzero, without any requirement on spacetime symmetry and charge distribution (Type II Paradox). We consider this issue as a fundamental flaw in Maxwell's equations.

We have investigated the possibilities for getting rid of the paradoxes by modifying Maxwell's equations. Although an electromagnetic field equation containing a photon mass term (i.e., the Proca equation) can address the problem, the origin of the photon mass term itself need be explained. Therefore we have considered an electromagnetic field equation with a curvature-coupled term, i.e., the field equation (\ref{meq2}). The new field equation can also address the paradoxes and has the following advantages over the Proca equation:

\begin{enumerate}

\item Unlike the Proca equation, the new field equation does not introduce a new dimensional parameter. In the Proca equation a photon mass is introduced as a fundamental constant. To be consistent with modern experiments and observations, the photon mass has to be incredibly small. In equation (\ref{meq2}), the $\kappa$ is a dimensionless constant of order unity.

\item In a flat or Ricci-flat spacetime, the new field equation returns to the Maxwell equation and gauge invariance is restored. In the Proca equation, the photon mass term is always there, and gauge invariance is always violated (though with introduction of a scalar field the gauge invariance can be recovered in the underlying theory). 

\item In a spacetime with a cosmological constant, the curvature-coupled term in the new field equation contains a photon mass term, where the photon mass is determined by the cosmological constant. In this sense, the new field equation implies the Proca equation and hence is more fundamental than the Proca equation.

\item The new field equation has a strong motivation from the ``minimal substitution rule'', as discussed in details in Sec.~\ref{der_ccm}. It can naturally arise from the Maxwell equation when the Maxwell equation is generalized to a curved spacetime with the ``minimal substitution rule''. The Proca equation has no similar motivation, unless the photon mass term is interpreted as arising from a more fundamental theory.

\item The Proca equation may never be testable, if the presumed photon mass is incredibly small. The new field equation contains a curvature-coupled term, whose effect is determined by the strength of the Ricci tensor. In a spacetime region that is strongly curved, the effect can be important and measurable. So, the new field equation (\ref{meq2}) is testable.

\end{enumerate}

We have discussed some consequences of the modified electromagnetic field equations and the testability of the new field equation (\ref{meq2}). Although in normal conditions the effect of a photon mass is hardly noticeable, a couple of places have been identified where the effect of the curvature-coupled term in the new field equation and the Lagrangian can be detectable. The places include the neighbor and interior of neutron stars, and the early universe. 

Due to the high mass density present inside a neutron star, a pseudo-charge current arising from the curvature-coupled term can affect the structure of the magnetic field of a neutron star. In the early time when the universe is small, the curvature-coupled term in the electromagnetic field equation affects all the electromagnetic processes on scales comparable to the curvature radius of the universe. The dark electromagnetic energy associated with the curvature-coupled term in the Lagrangian may also play an important role in driving the expansion of the universe.

In summary, to address the inconsistency problem arising from application of the Maxwell equation to a universe with a net charge, we have proposed and investigated a new electromagnetic field equation (eq.~\ref{meq2}). The new field equation is strongly motivated by consideration of generalization of the Maxwell equation in a flat spacetime to a curved spacetime, and is testable in a strongly curved spacetime region and in the early epoch of the universe.

\begin{acknowledgments}
The author thanks an anonymous reviewer for a very good report which has helped to improve the presentation of the paper. This work was supported by the National Basic Research Program (973 Program) of China (Grant No.~2014CB845800) and the NSFC grants program (no.~11373012).
\end{acknowledgments}

\appendix

\section{Derivation of Equation (\ref{Tab2})}
\label{Tabs}

By the equation 7.5.14 of \cite{wal84}, we have
\begin{eqnarray}
        2\delta R_{ac} = -g^{bd}\nabla_a\nabla_c\delta g_{bd}-g^{bd}\nabla_b\nabla_d\delta g_{ac} +2g^{bd}\nabla_b\nabla_{(a}\delta g_{c)d} \;,
\end{eqnarray}
and
\begin{eqnarray}
        2A^aA^c\delta R_{ac} = -g^{bd}A^aA^c\nabla_a\nabla_c\delta g_{bd}-g^{bd}A^aA^c\nabla_b\nabla_d\delta g_{ac} +2g^{bd}A^aA^c\nabla_b\nabla_a\delta g_{cd} \;. \label{dR1}
\end{eqnarray}

Let us evaluate each term on the right-hand side of equation (\ref{dR1}) separately. For the first term, we have
\begin{eqnarray}
        -g^{bd}A^aA^c\nabla_a\nabla_c\delta g_{bd} &=& -g^{bd}\nabla_a\left(A^aA^c\nabla_c\delta g_{bd}\right)+g^{bd}\nabla_a(A^aA^c)\nabla_c\delta g_{bd} \nonumber\\
        &=& -g^{bd}\nabla_a\left(A^aA^c\nabla_c\delta g_{bd}\right)+g^{bd}\nabla_c\left[\nabla_a(A^aA^c)\delta g_{bd}\right] -g^{bd}\nabla_c\nabla_a(A^aA^c)\delta g_{bd} \nonumber\\
        &=& \nabla^a[...]_a-g^{bd}\nabla_c\nabla_a(A^aA^c)\delta g_{bd} \;,
\end{eqnarray}
where $\nabla^a[...]_a$ is not written out explicitly since it does not contribute to the integral of action so does not affect the derivation of stress-energy tensor. 

Similarly, for the second term on the right-hand side of equation (\ref{dR1}), we have
\begin{eqnarray}
        -g^{bd}A^aA^c\nabla_b\nabla_d\delta g_{ac} = \nabla^a[...]_a-g^{bd}\nabla_d\nabla_b(A^aA^c)\delta g_{ac} \;.
\end{eqnarray}
For the third term, we have
\begin{eqnarray}
        +2g^{bd}A^aA^c\nabla_b\nabla_a\delta g_{cd} =\nabla^a[...]_a+2g^{bd}\nabla_a\nabla_b(A^aA^c)\delta g_{cd} \;.
\end{eqnarray}

Hence we get
\begin{eqnarray}
        2A^aA^c\delta R_{ac} \doteq -g^{bd}\nabla_c\nabla_a(A^aA^c)\delta g_{bd} -g^{bd}\nabla_d\nabla_b(A^aA^c)\delta g_{ac}+2g^{bd}\nabla_a\nabla_b(A^aA^c)\delta g_{cd} \;, 
\end{eqnarray}
where $\doteq$ means ``equal up to a term like $\nabla^a[...]_a$''. Making use of the identity $\delta g_{ab}=-g_{ac}g_{bd}\delta g^{cd}$, we get
\begin{eqnarray}
        2A^aA^c\delta R_{ac} \doteq \left[\nabla_c\nabla_d(A^cA^d)g_{ab} +\nabla^d\nabla_d(A_aA_b)-2\nabla^c\nabla_b(A_aA_c)\right]\delta g^{ab} \;.
\end{eqnarray}

On the other hand, we have
\begin{eqnarray}
        R_{ac}\delta(A^aA^c) = 2A^cR_{ca}A_b\delta g^{ab} \;,
\end{eqnarray}
and
\begin{eqnarray}
        R_{ac}A^aA^c\delta\sqrt{-g}=-\frac{1}{2}\sqrt{-g}R_{cd}A^cA^dg_{ab}\delta g^{ab} \;. 
\end{eqnarray}

Therefore, we have
\begin{eqnarray}
        \delta\left(\sqrt{-g}R_{ac}A^aA^c\right) \doteq \sqrt{-g}\left\{\frac{1}{2}\nabla^c\nabla_c(A_aA_b)-\nabla^c\nabla_b(A_cA_a)+2A^cR_{ca}A_b+\frac{1}{2}g_{ab}\left[\nabla_c\nabla_d(A^cA^d) -R_{cd}A^cA^d\right]\right\} \delta g^{ab} \;. \nonumber\\ 
\end{eqnarray}

Then, by equation (\ref{T_em_ab}), the stress-energy tensor in equation (\ref{Tab2}) is derived after symmetrization of the tensor index as required by the definition of $T_{ab}$.

\section{Derivation of Equation (\ref{dTab2})}
\label{dTabs}

For the simplicity of calculations, let us define
\begin{eqnarray}
        \Pi_{ab}^{(1)} = \nabla_c\nabla_d(A^cA^d)g_{ab} +\nabla^c\nabla_c(A_aA_b) -2\nabla^c\nabla_{(a}(A_{b)}A_c) \;, \label{Pi(1)}
\end{eqnarray}
and
\begin{eqnarray}
        \Pi_{ab}^{(2)} = 4A^cR_{c(a}A_{b)}-R_{cd}A^cA^dg_{ab} \;. \label{Pi(2)}
\end{eqnarray}
We have then
\begin{eqnarray}
        T_{\emm,ab} = ~^{(0)}T_{\emm,ab}+\frac{\kappa}{8\pi}\left(\Pi_{ab}^{(1)}+\Pi_{ab}^{(2)}\right) \;, \label{Tab2s}
\end{eqnarray}
and
\begin{eqnarray}
        \nabla^aT_{\emm,ab} = \nabla^a\!\!~^{(0)}T_{\emm,ab}+\frac{\kappa}{8\pi}\nabla^a\left(\Pi_{ab}^{(1)}+\Pi_{ab}^{(2)}\right) \;, \label{dTab2s}
\end{eqnarray}
where
\begin{eqnarray}
        \nabla^a\!\!~^{(0)}T_{\emm,ab} = \frac{1}{4\pi}F_{ba}\nabla_cF^{ca} \;. \label{dT0}
\end{eqnarray}

The divergence of $\Pi_{ab}^{(1)}$ is
\begin{eqnarray}
        &&\nabla^a\Pi_{ab}^{(1)} = \nabla_b\nabla_c\nabla_d(A^cA^d) +\nabla^a\nabla^d\nabla_d(A_aA_b)-2\nabla^a\nabla^c\nabla_{(a}(A_{b)}A_c) \nonumber\\
        &&~= \nabla_b\left(\nabla_c A^c\nabla_dA^d+A^c\nabla_c\nabla_dA^d +\nabla_cA^d\nabla_dA^c+A^d\nabla_c\nabla_dA^c\right) +\nabla^a\left(2\nabla^dA_a\nabla_dA_b+A_a\nabla^d\nabla_dA_b +A_b\nabla^d\nabla_dA_a\right) \nonumber\\
        &&~\hspace{4mm}-\nabla^a\left(\nabla^cA_b\nabla_aA_c+A_b\nabla^c\nabla_aA_c +\nabla^cA_c\nabla_aA_b+A_c\nabla^c\nabla_aA_b\right)-\nabla^a\left(\nabla^cA_a\nabla_bA_c+A_a\nabla^c\nabla_bA_c +\nabla^cA_c\nabla_bA_a\right. \nonumber\\
        &&~\hspace{4mm}\left.+A_c\nabla^c\nabla_bA_a\right) \nonumber\\
        &&~= \left(2\nabla_b\nabla_c A^c\nabla_dA^d +\nabla_bA^c\nabla_c\nabla_dA^d+A^c\nabla_b\nabla_c\nabla_dA^d+2\nabla_b\nabla_cA^d\nabla_dA^c +\nabla_bA^d\nabla_c\nabla_dA^c+A^d\nabla_b\nabla_c\nabla_dA^c\right) \nonumber\\
        &&~\hspace{4mm}+\left(2\nabla^a\nabla^dA_a\nabla_dA_b+2\nabla^dA_a\nabla^a\nabla_dA_b +\nabla^aA_a\nabla^d\nabla_dA_b+A_a\nabla^a\nabla^d\nabla_dA_b+\nabla^aA_b\nabla^d\nabla_dA_a+A_b\nabla^a\nabla^d\nabla_dA_a\right) \nonumber\\
        &&~\hspace{4mm}-\left(\nabla^a\nabla^cA_b\nabla_aA_c+\nabla^cA_b\nabla^a\nabla_aA_c +\nabla^aA_b\nabla^c\nabla_aA_c+A_b\nabla^a\nabla^c\nabla_aA_c+\nabla^a\nabla^cA_c\nabla_aA_b+\nabla^cA_c\nabla^a\nabla_aA_b \right.\nonumber\\
        &&~\hspace{4mm}\left.+\nabla^aA_c\nabla^c\nabla_aA_b+A_c\nabla^a\nabla^c\nabla_aA_b\right)-\left(\nabla^a\nabla^cA_a\nabla_bA_c+\nabla^cA_a\nabla^a\nabla_bA_c +\nabla^aA_a\nabla^c\nabla_bA_c+A_a\nabla^a\nabla^c\nabla_bA_c\right.\nonumber\\
        &&~\hspace{4mm}\left.+\nabla^a\nabla^cA_c\nabla_bA_a+\nabla^cA_c\nabla^a\nabla_bA_a +\nabla^aA_c\nabla^c\nabla_bA_a+A_c\nabla^a\nabla^c\nabla_bA_a\right) \nonumber\\
        &&~=\nabla_aA^a\times\circled{1}+\nabla_bA^c\times\circled{2}+\nabla^cA_b\times\circled{3}+\nabla_aA^c\times\circled{4}+[\mbox{Rest}]\;,
\end{eqnarray}
where
\begin{eqnarray}
        \circled{1} = 2\left(\nabla_b\nabla_c-\nabla_c\nabla_b\right) A^c =-2R_{bd}A^d\;, \hspace{1cm} \circled{2} =0 \;, \hspace{1cm} \circled{3} = (\nabla_a\nabla_c-\nabla_c\nabla_a)A^a =R_{cd}A^d \;, \\[3mm]
        \circled{4} = 2(\nabla_b\nabla_c-\nabla_c\nabla_b)A^a+\left(\nabla_c\nabla^a-\nabla^a\nabla_c\right)A_b=-2R_{bcd}^{\;\;\;\;\;\;a}A^d+R_{bdc}^{\;\;\;\;\;\;a}A^d \;, \hspace{1.6cm}~
\end{eqnarray}
and
\begin{eqnarray}
        [\mbox{Rest}] &=& A^c\nabla_b\nabla_c\nabla_dA^d +A^d\nabla_b\nabla_c\nabla_dA^c +A_a\nabla^a\nabla^d\nabla_dA_b +A_b\nabla^a\nabla^d\nabla_dA_a -A_b\nabla^a\nabla^c\nabla_aA_c -A_c\nabla^a\nabla^c\nabla_aA_b \nonumber\\
        && -A_a\nabla^a\nabla^c\nabla_bA_c -A_c\nabla^a\nabla^c\nabla_bA_a \nonumber\\
        &=& A^c\left[\nabla_b\left(\nabla_c\nabla_aA^a+\nabla_a\nabla_cA^a\right)-\left(\nabla_c\nabla_a+\nabla_a\nabla_c\right)\nabla_bA^a\right]+A_b\left(\nabla_a\nabla_c-\nabla_c\nabla_a\right)\nabla^cA^a \nonumber\\
        &&+A^a\left(\nabla_a\nabla_c-\nabla_c\nabla_a\right)\nabla^cA_b \;.
\end{eqnarray}

By the equation 3.2.12 of \cite{wal84}, we have $\nabla_a\nabla_b\omega_{cd}-\nabla_b\nabla_a\omega_{cd}=R_{abc}^{\;\;\;\;\;\;e}\omega_{ed}+R_{abd}^{\;\;\;\;\;\;e}\omega_{ce}$ for any tensor $\omega_{ab}$, from which we get $\nabla_a\nabla_b\omega^c_{\;\,d}-\nabla_b\nabla_a\omega^c_{\;\,d}=-R_{abe}^{\;\;\;\;\;\;c}\omega^e_{\;\,d}+R_{abd}^{\;\;\;\;\;\;e}\omega^c_{\;\,e}$. Hence, we have
\begin{eqnarray}
        \left(\nabla_c\nabla_a+\nabla_a\nabla_c\right)\nabla_bA^a = 2\nabla_c\nabla_a\nabla_bA^a+R_{acb}^{\;\;\;\;\;\;e}\nabla_eA^a+R_{ce}\nabla_bA^e \;,
\end{eqnarray}
where we have used the identities $R_{ace}^{\;\;\;\;\;\;a}=-R_{cae}^{\;\;\;\;\;\;a}=-R_{ce}$.

We also have $\nabla_c\nabla_aA^a+\nabla_a\nabla_cA^a=2\nabla_c\nabla_aA^a+R_{cd}A^d$. Then, in the expression for [Rest],
\begin{eqnarray}
        A^c\left[\nabla_b\left(\nabla_c\nabla_aA^a+\nabla_a\nabla_cA^a\right)-\left(\nabla_c\nabla_a+\nabla_a\nabla_c\right)\nabla_bA^a\right] =A^c\left[2\nabla_c\left(\nabla_b\nabla_aA^a-\nabla_a\nabla_bA^a\right)+\nabla_b\left(R_{cd}A^d\right)\right. \nonumber\\
        \left.-R_{acb}^{\;\;\;\;\;\;e}\nabla_eA^a-R_{ce}\nabla_bA^e\right] =A^c\left[-2\nabla_c\left(R_{bd}A^d\right)+\nabla_b\left(R_{cd}A^d\right)-R_{acb}^{\;\;\;\;\;\;e}\nabla_eA^a-R_{ce}\nabla_bA^e\right] \;, \hspace{0.8cm}~ 
\end{eqnarray}
where we have used the identity $\nabla_b\nabla_c\nabla_aA^a=\nabla_c\nabla_b\nabla_aA^a$.

For the other two terms in [Rest], we have
\begin{eqnarray}
        A_b\left(\nabla_a\nabla_c-\nabla_c\nabla_a\right)\nabla^cA^a = A_b\left(-R_{ace}^{\;\;\;\;\;\;c}\nabla^eA^a-R_{ace}^{\;\;\;\;\;\;a}\nabla^cA^e\right) = -A_bR_{ae}\left(\nabla^eA^a-\nabla^aA^e\right) = 0 \;,
\end{eqnarray}
and
\begin{eqnarray}
        A^a\left(\nabla_a\nabla_c-\nabla_c\nabla_a\right)\nabla^cA_b = A^a\left(-R_{ae}\nabla^eA_b +R_{acb}^{\;\;\;\;\;\;e}\nabla^cA_e\right) \;.
\end{eqnarray}
Therefore, we have
\begin{eqnarray}
        [\mbox{Rest}] &=& A^c\left[-2\nabla_c\left(R_{bd}A^d\right)+\nabla_b\left(R_{cd}A^d\right)-R_{acb}^{\;\;\;\;\;\;e}\nabla_eA^a-R_{ce}\nabla_bA^e\right]+A^a\left(-R_{ae}\nabla^eA_b+R_{acb}^{\;\;\;\;\;\;e}\nabla^cA_e\right) \nonumber\\
        &=& A^c\left[-2\nabla_c\left(R_{bd}A^d\right)+\nabla_b\left(R_{cd}A^d\right)+R_{cabe}(\nabla^eA^a+\nabla^aA^e)-R_{ce}\left(\nabla_bA^e+\nabla^eA_b\right)\right] \;. 
\end{eqnarray}

Putting the above results together, we get
\begin{eqnarray}
        \nabla^a\Pi_{ab}^{(1)} = -2\nabla_c\left(A^cR_{bd}A^d\right)+A^c\nabla_b\left(R_{cd}A^d\right)-A^dR_{dc}\nabla_bA^c \;, \label{dPi(1)}
\end{eqnarray}
where we have used the identity $-R_{bcda}+R_{bdca}+R_{dcba}=0$ (since $R_{[abc]}^{\;\;\;\;\;\;\;d}=0$). 

The divergence of $\Pi_{ab}^{(2)}$ is
\begin{eqnarray}
        \nabla^a\Pi_{ab}^{(2)} = 2\nabla_c\left(A^dR_{db}A^{c}\right)+2\nabla^c\left(A^dR_{dc}A_{b}\right) -\nabla_b\left(R_{cd}A^cA^d\right) \;. \label{dPi(2)}
\end{eqnarray}
Therefore, we have
\begin{eqnarray}
        \nabla^a\left(\Pi_{ab}^{(1)}+\Pi_{ab}^{(2)}\right) = 2\nabla^c\left(A^dR_{dc}A_{b}\right)-2A^dR_{dc}\nabla_bA^c = -2F_b^{\;\,c}R_{cd}A^d+2A_b\nabla^c\left(R_{cd}A^d\right) \;. \label{dPi}
\end{eqnarray}

Substituting equations (\ref{dT0}) and (\ref{dPi}) into equation (\ref{dTab2s}), we get equation (\ref{dTab2}).

Note, when we derived the equation (\ref{dPi}) we did not make use of the electromagnetic field equation. When $R_{ab}=0$, we have $\Pi_{ab}^{(2)}=0$ and $\nabla^a\Pi_{ab}^{(1)}=0$.

\section{Derivation of the Poynting Flux Vector}
\label{EB}

In this appendix we derive the Poynting flux vector corresponding to the electromagnetic field equation (\ref{meq1}).

By equations (\ref{Ea}) and (\ref{Ba}), the $F_{ab}$ can be expressed in terms of $E_a$ and $B_a$ as
\begin{eqnarray}
	F_{ab}=-2E_{[a}u_{b]}-\epsilon_{abcd}B^cu^d \;, \label{F_E_B}
\end{eqnarray}
where $u^au_a=-1$ and $E_au^a=B_au^a=0$. We can derive
\begin{eqnarray}
	F_{ab}F^{ab}=-2\left(E_aE^a-B_aB^a\right) \;,
\end{eqnarray}
and
\begin{eqnarray}
	F_{ac}F_b^{\;\,c} = -E_aE_b-B_aB_b+E_cE^cu_au_b +(g_{ab}+u_au_b)B_cB^c+2u_{(a}\tilde{\epsilon}_{b)cd}E^cB^d \;, \label{Tab_de1}
\end{eqnarray}
where $\tilde{\epsilon}_{abc}\equiv -\epsilon_{abcd}u^d$.

By equation (\ref{0TEMab}), we get
\begin{eqnarray}
	4\pi\!\!~^{(0)}T_{\emm,ab} = \left(E_cE^c+B_cB^c\right)\left(u_au_b+\frac{1}{2}g_{ab}\right) -E_aE_b-B_aB_b+2u_{(a}\tilde{\epsilon}_{b)cd}E^cB^d \;. \label{Tab_EB}
\end{eqnarray}

To express the law of energy conservation in terms of $E_a$ and $B_a$, we assume that the spacetime has a timelike Killing vector $t^a$ parallel to the $u^a$, i.e., $t^a=\alpha u^a$, where $\alpha$ is the lapse function. By the Killing equation \cite{wal84}, we get
\begin{eqnarray}
	0=\nabla_{(a}t_{b)}=\alpha\nabla_{(a}u_{b)}+u_{(a}\nabla_{b)}\alpha \;,
\end{eqnarray}
from which we can derive
\begin{eqnarray}
	\alpha\nabla_au^a+u^a\nabla_a\alpha =0 \label{kill1}
\end{eqnarray}
and
\begin{eqnarray}
        u^a\nabla_au_b=\nabla_b\ln\alpha-u_bu^a\nabla_a\ln\alpha \;. \label{kill2}
\end{eqnarray}
Since $u^bu_b=-1$ and $u^b\nabla_au_b=0$, equation (\ref{kill2}) leads to
\begin{eqnarray}
	u^a\nabla_a\alpha=0 \;, \hspace{1cm} u^a\nabla_au_b=\nabla_b\ln\alpha \;. \label{kill4}
\end{eqnarray}
Then, by equation (\ref{kill1}) we have
\begin{eqnarray}
	\nabla_au^a=0 \;. \label{kill5}
\end{eqnarray}

An energy-momentum flux vector ${\cal J}^a$ is defined by
\begin{eqnarray}
	{\cal J}_a \equiv -t^b T_{\emm,ab} = -\alpha u^b T_{\emm,ab}\;. \label{cJ0}
\end{eqnarray}
By equations (\ref{TEMab}), (\ref{Tab_EB}), and (\ref{1TEMab}), we get
\begin{eqnarray}
	{\cal J}^a = \frac{\alpha}{4\pi}\left\{\frac{1}{2}\left[E_cE^c+B_cB^c+\xi\left(\Phi^2+\tilde{A}_c\tilde{A}^c\right)\right]u^a+\left(\tilde{\epsilon}^{acd}E_cB_d+\xi\Phi\tilde{A}^a\right)\right\} \;, \label{cJa}
\end{eqnarray}
where we have written $A^a$ in terms of a scalar potential $\Phi$ and a spatial vector potential $\tilde{A}^a$ (i.e., $A^a=\Phi u^a+\tilde{A}^a$, $u_a\tilde{A}^a=0$).

With the expression in equation (\ref{cJa}), the energy-momentum flux vector is decomposed into an energy density component parallel to $u^a$, and a spatial momentum component perpendicular to $u^a$:
\begin{eqnarray}
	{\cal J}^a=\rho_\emm u^a+\tilde{\cal J}^a \;, \label{cJa_decom}
\end{eqnarray}
where
\begin{eqnarray}
	\rho_\emm=\frac{\alpha}{8\pi}\left[E_cE^c+B_cB^c+\xi\left(\Phi^2+\tilde{A}_c\tilde{A}^c\right)\right] \;, \label{rho_em}
\end{eqnarray}
and
\begin{eqnarray}
	\tilde{\cal J}^a=\frac{\alpha}{4\pi}\left(\tilde{\epsilon}^{acd}E_cB_d+\xi\Phi\tilde{A}^a\right) \;. \label{cJ_mom}
\end{eqnarray}

In $\tilde{\cal J}^a$, the first term is just the usual Poynting flux vector, and the second term is an additional momentum flux due to the $\xi$-term in the electromagnetic equation. We may call $\tilde{\cal J}^a$ the generalized Poynting flux vector.

\end{document}